\documentclass[acmlarge]{acmart}

\usepackage{multirow}
\usepackage{enumitem}
\usepackage{pifont}
\usepackage{makecell}
\usepackage[table]{xcolor}
\usepackage{graphicx}
\usepackage{subcaption}
\usepackage{booktabs} 
\usepackage{tabularx} 
\usepackage{array}    
\usepackage{seqsplit} 
\usepackage{listings}
\usepackage{lineno}
\usepackage{times}
\usepackage{latexsym}
\usepackage{xltabular}
\usepackage{stfloats}
\usepackage{placeins}
\usepackage[T1]{fontenc}
\usepackage[utf8]{inputenc}
\usepackage{microtype}
\usepackage{graphicx}
\usepackage{amsmath}
 
\usepackage{amssymb}
 \usepackage{booktabs} 
\usepackage{pifont}     
\usepackage{colortbl}   
\usepackage{tabularx}
\usepackage[table]{xcolor}
\usepackage{seqsplit}
\usepackage{multicol}
\usepackage{enumitem}
\usepackage{multirow}
\usepackage{caption}
\usepackage{longtable}
\usepackage[most]{tcolorbox}
\usepackage{array}
\usepackage{ragged2e} 
\usepackage{eurosym}

\newcolumntype{L}{>{\raggedright\arraybackslash}X}
\newcolumntype{I}{>{\centering\arraybackslash}X}
\newcolumntype{S}{>{\raggedright\arraybackslash}X}
\definecolor{lightgray}{gray}{0.9}

\newcommand{\framework}{\textsc{FinDeepForecast}}
\newcommand{\bench}{\textsc{FinDeepForecastBench}}

\newcommand{\ie}{\emph{i.e., }}
\newcommand{\eg}{\emph{e.g., }}

\newcommand{\headernodot}[1]{\noindent\textbf{#1}}

\usepackage{subcaption}

\begin{document}

\title{\framework~: A Live Multi-Agent System for Benchmarking Deep Research Agents in Financial Forecasting}

\subtitle{\textbf{Project Website}: \textcolor{blue}{\textbf{\url{https://OpenFinArena.com/}}}}

\author{Xiangyu Li$^{\ast\clubsuit\dagger}$, Xuan Yao$^{\ast\spadesuit}$, Guohao Qi$^{\ast\clubsuit}$, Fengbin Zhu$^{\ast\ddagger\clubsuit}$, Kelvin J.L. Koa$^{\spadesuit}$, Xiang Yao Ng$^{\heartsuit}$, Ziyang Liu$^{\heartsuit}$, Xingyu Ni$^{\spadesuit}$, Chang Liu$^{\spadesuit}$, Yonghui Yang$^{\clubsuit}$, Yang Zhang$^{\clubsuit}$, Wenjie Wang$^{\diamondsuit}$, Fuli Feng$^{\diamondsuit}$, Chao Wang$^{\heartsuit}$, Huanbo Luan$^{\heartsuit}$, Xiaofen Xing$^{\dagger}$, Xiangmin Xu$^{\dagger}$, Tat-Seng Chua$^{\clubsuit}$, Ke-Wei Huang$^{\spadesuit}$}

\affiliation{
\institution{ \\
$^{\clubsuit}$National University of Singapore, Singapore\\
$^{\spadesuit}$Asian Institute of Digital Finance, Singapore\\
$^{\heartsuit}$6Estates Pte Ltd, Singapore\\
$^{\diamondsuit}$University of Science and Technology of China, China\\
$^{\dagger}$South China University of Technology}\country{China}
}

\thanks{$^{\ast}$\textbf{Equal Contribution}}

\thanks{$^{\ddagger}$\textbf{Project Owner \& Corresponding Author: Fengbin Zhu, fengbin@nus.edu.sg}}

\renewcommand{\shortauthors}{Xiangyu Li, et al}

\begin{abstract}
Deep Research (DR) Agents powered by advanced Large Language Models (LLMs) have fundamentally shifted the paradigm for completing complex research tasks.
Yet, a comprehensive and live evaluation of their forecasting performance on real-world, research-oriented tasks in high-stakes domains (\eg finance) remains underexplored.
We introduce \framework, the first live, end-to-end multi-agent system for automatically evaluating DR agents by continuously generating research-oriented financial forecasting tasks.
This system is equipped with a \emph{dual-track taxonomy}, enabling the dynamic generation of recurrent and non-recurrent forecasting tasks at both corporate and macro levels.
With this system, we generate \bench, a weekly evaluation benchmark over a ten-week horizon, encompassing 8 global economies and 1,314 listed companies, and evaluate 13 representative methods.
Extensive experiments show that, while DR agents consistently outperform strong baselines, their performance still falls short of genuine forward-looking financial reasoning.
We expect the proposed \framework~system to consistently facilitate future advancements of DR agents in research-oriented financial forecasting tasks. The benchmark and leaderboard are publicly available on the \href{https://OpenFinArena.com/}{OpenFinArena Platform}. 
\end{abstract}



\maketitle

\section{Introduction}
Deep Research (DR) agents are autonomous artificial intelligence (AI) systems that perform complex research tasks via iterative planning, evidence acquisition, reasoning, and reporting~\cite{shi2025deep,huang2025survey}. 
Their emergence has reshaped the approach to complex tasks, attracting growing research attention.
Beyond developing more powerful DR agents, reliably evaluating their research capabilities is an equally fundamental objective for advancing the field, yet is currently underexplored.
Traditional static evaluation benchmarks inevitably leak into training corpora, rendering them obsolete (\ie data contamination)~\cite{xu2024contamination,balloccu2024leak}. 
Recently, live benchmarks have been explored, continuously generating novel instances to ensure temporal separation between training and evaluation data~\cite{white2025livebench}.

\begin{figure}[t]
\centering
\includegraphics[width=0.999\textwidth]{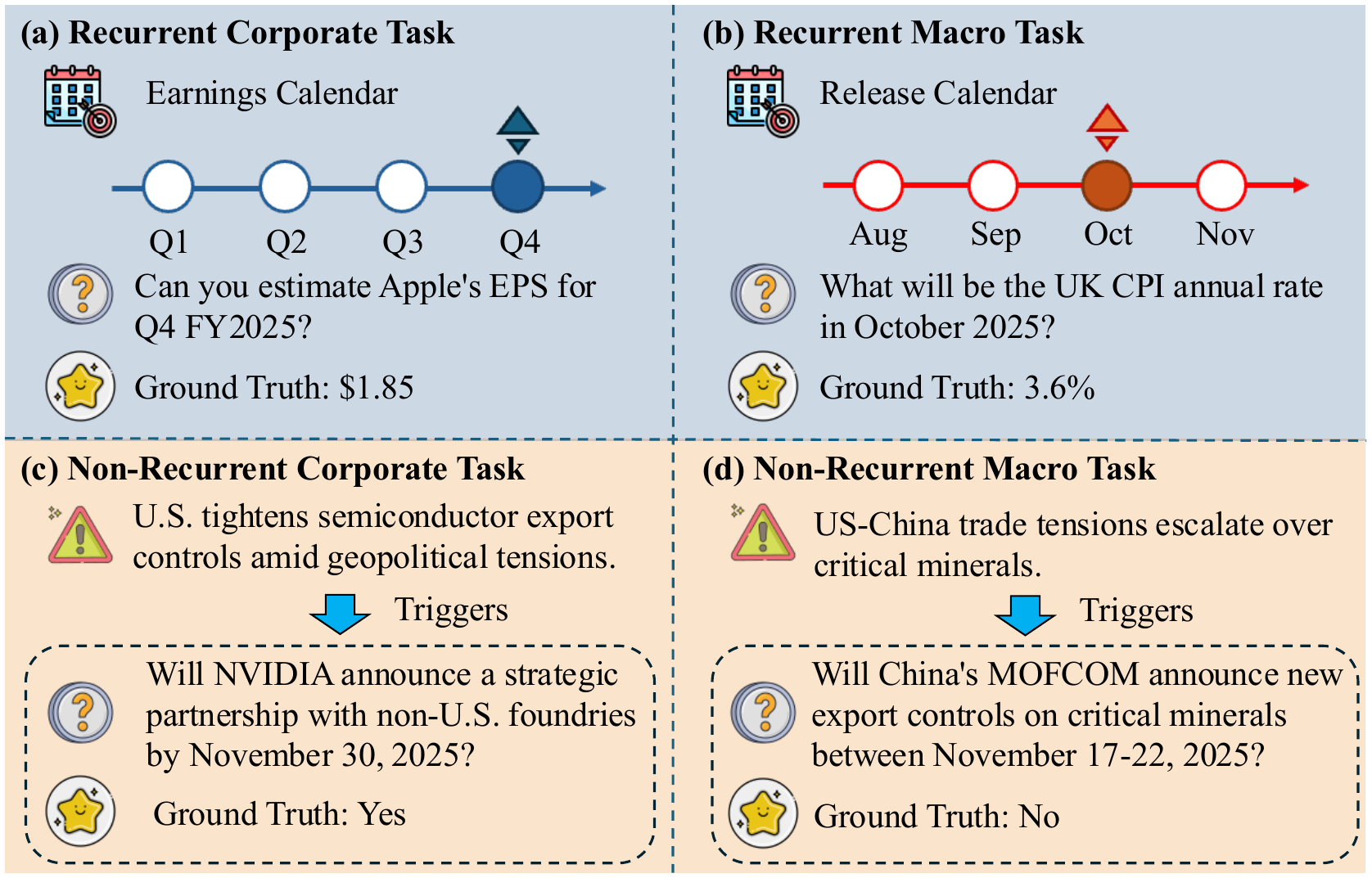}
\caption{Recurrent tasks for regular disclosures and non-recurrent tasks for event-driven predictions.}
\vspace{-0.6cm}
\label{fig:taxonomy}
\end{figure}

Existing live benchmarks are either \emph{time-insensitive} or \emph{time-sensitive}.
The former focus on domains with static and deterministic ground-truths, like code generation~\cite{jain2024livecodebench,zhang2025swe} and mathematics problem-solving~\cite{balunovic2025matharena}, often leading to insufficiently rigorous evaluations---models may rely on the recall of pre-existing answer patterns rather than genuine reasoning.
Time-sensitive benchmarks~\cite{karger2024forecastbench,zeng2025futurex} are constructed around future data, undisclosed content, or unknown outcomes, such that correct answers cannot be known before evaluation, enabling rigorous assessment of predictive reasoning.
However, the tasks in these benchmarks are often sampled from existing question websites or generated using fixed templates. 
Such heavy reliance on external sources or predefined templates can introduce inherent biases and limit the availability of genuinely research-oriented tasks, thereby constraining the breadth and depth of the evaluation. 
To truly benchmark the capabilities of DR agents, a dynamic evaluation environment is needed—one that continuously supplies \textbf{forward-looking, research-oriented} tasks with strict temporal constraints and objectively verifiable outcomes.

The financial domain offers an ideal setting for such forward-looking, research-oriented tasks~\cite{koval2024financial, chen2024knowledge}, with properties that support the ongoing evaluation of DR agents’ forecasting capabilities.
1) It offers \emph{Periodic Information Disclosure}, with a continuous stream of verifiable data points,
such as corporate financial metrics (\eg EPS in Fig.~\ref{fig:taxonomy} (a)) and macroeconomic (macro) indicators (\eg CPI in Fig.~\ref{fig:taxonomy} (b)).
2) It features \emph{Diverse Market Events}, necessitating the distillation of valuable signals and market dynamics from vast, noisy data for accurately forecasting various critical events, including corporate actions (\eg new partnership in Fig.~\ref{fig:taxonomy} (c)) and macro shifts (\eg export control in  Fig.~\ref{fig:taxonomy} (d)).
3) It guarantees \emph{Strict Temporal Isolation}, with answers emerging strictly upon disclosure of information or occurrence of the corresponding events.
In such financial markets, financial experts have to gather and analyze information and forecast future outcomes through reasoning in order to complete complex tasks as shown in Fig.~\ref{fig:taxonomy}.

Aimed at continuously evaluating DR agents in addressing research-oriented financial forecasting challenges, we propose a novel, live multi-agent system, named \framework.
As shown in Fig.~\ref{fig:framework},  it employs a \emph{dual-track taxonomy} (see Appendix~\ref{app:recurrent} and ~\ref{app:nonrecurrent} for more details) for effectively managing recurrent and non-recurrent forecasting scenarios, encompassing corporate- and macro-level tasks.
It comprises four key stages, powered by six specialized agents, for an automatic, end-to-end evaluation of DR agents, starting from data collection and task generation through model forecasting to ground-truth acquisition and performance evaluation.
To validate this system, we generate a \bench~benchmark, which covers 8 major global economies for macro tasks and 1,314 listed companies drawn from 9 major indices for corporate tasks.
In total, it consists of 1,394 tasks, including 296 recurrent macro, 723 recurrent corporate, 128 non-recurrent macro, and 247 non-recurrent corporate tasks.

We assess 13 representative methods for completing the weekly tasks in \bench, including 3 DR agents, 5 LLMs with both thinking and search capabilities, and 5 LLMs with thinking capabilities.
Several important findings have been made. 
1) DR agents consistently exhibit superior performance to the compared methods, but they still struggle significantly with the tasks in \bench, with the highest score 39.5 out of 100.
2) Most methods achieve peak performance in information-rich markets (\eg US and China) but underperform in markets with relatively limited data or language diversity (\eg Japan).
3) The models achieve high accuracy on non-recurrent tasks, but their performance declines sharply on recurrent tasks. 
This disparity underscores the greater intrinsic difficulty of precise numeric forecasting under periodic disclosure when compared to binary event prediction.

\begin{figure*}[t]
\centering
\includegraphics[width=0.999\textwidth]{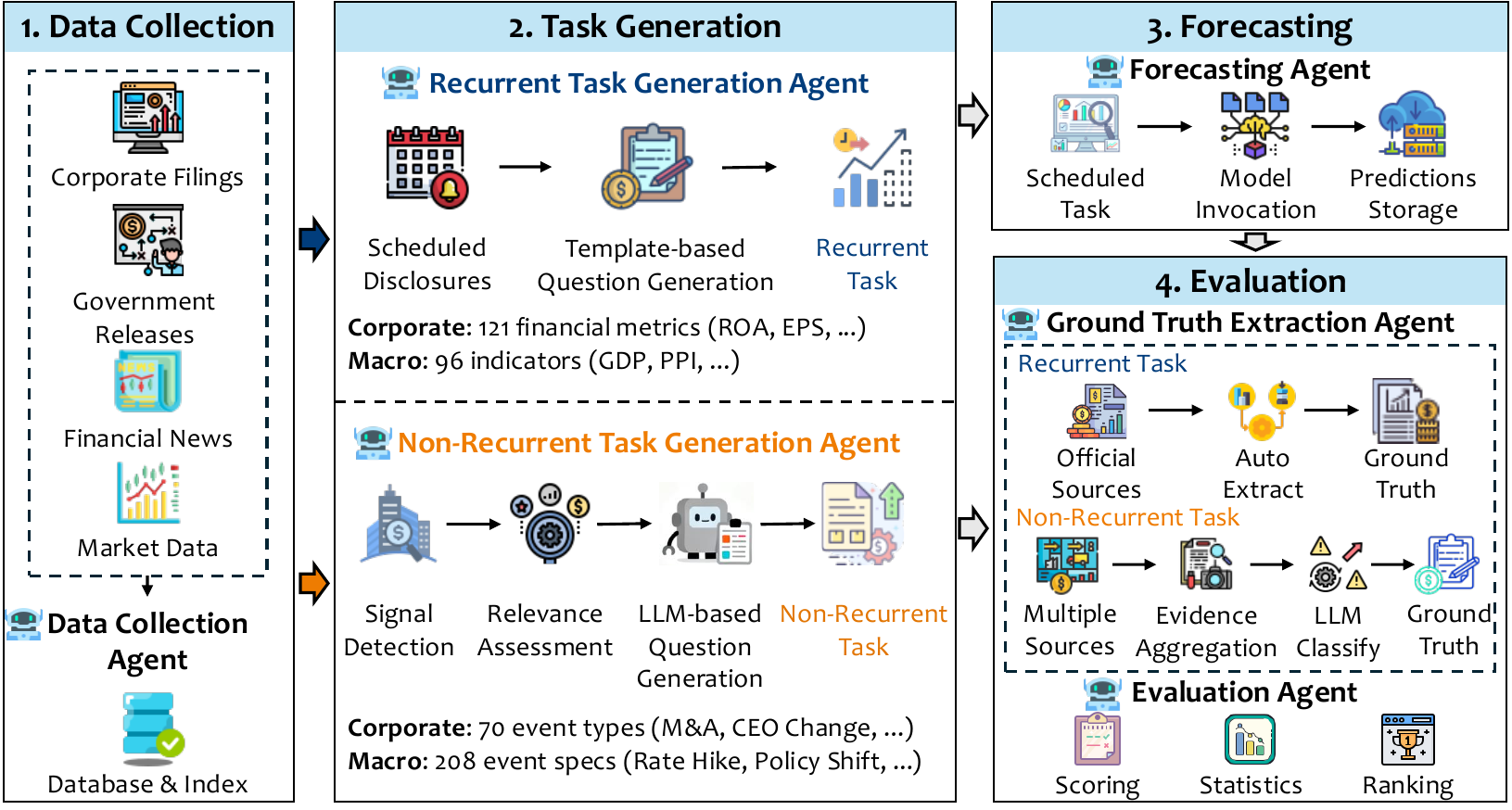}
\caption{The \framework{} system comprises four stages: (1) Data Collection aggregates financial information into a timestamped database; (2) Task Generation produces \textcolor{blue}{recurrent tasks} via template-based question generation and \textcolor{orange}{non-recurrent tasks} via LLM-based pipeline; (3) Forecasting invokes models with temporal isolation; (4) Evaluation extracts ground truth and computes performance metrics.}
\vspace{-0.6cm}
\label{fig:framework}
\end{figure*}

Our contributions are summarized as follows:
\begin{itemize}[leftmargin=*, itemsep=0pt, topsep=0pt]
    \item We develop \framework, \textbf{the first end-to-end multi-agent system} designed to continuously produce \textbf{forward-looking, research-oriented tasks} in finance for the contamination-free evaluation of DR agents.
    \item We propose a \textbf{dual-track taxonomy} for the dynamic generation of both recurrent and non-recurrent financial forecasting tasks, encompassing corporate- and macro-level predictions (covering hundreds of metrics and event categories) within a live market environment.
    \item With \framework, we generate \bench, a \textbf{weekly evaluation benchmark} spanning a ten-week horizon, currently covering 8 major global economies and 1,314 listed companies from 9 leading indices, while remaining readily extensible to additional markets and firms.
    \item  Extensive evaluations of 13 representative methods show that, although DR agents significantly outperform alternative approaches, they still exhibit substantial room for improvement, highlighting the limitations of current methods in solving these tasks.
    This firmly establishes our \framework~as a timely and essential contribution for consistently facilitating the future advancement of DR agents.
\end{itemize}

\section{\framework~System}

In this section, we introduce \framework{}, a live, multi-agent system for assessing genuine capabilities of DR agents in financial forecasting through research-oriented task generation, strict temporal isolation, and rigorous ground truth verification, as shown in Fig.~\ref{fig:framework}.
In \framework, a \textit{dual-track taxonomy} is devised to distinguish recurrent predictions for numerical estimation on scheduled disclosures from non-recurrent predictions for binary classification on uncertain emerging events. 
\textit{Continuous generation with temporal isolation} prevents data contamination through live task creation while enforcing uniform information boundaries across models.

\subsection{Forecasting Problem Definition}

A forecasting problem is defined as a tuple $\mathcal{P} = (q, t_g, t_d, t_e, y)$, where $q$ denotes the forecasting question, $t_g$ the task generation time, $t_d$ the forecasting deadline, $t_e$ the evaluation time, and $y$ the ground truth outcome. The temporal ordering $t_g < t_d < t_e$ ensures forecasts are made before outcomes become observable. Given a forecasting problem, a model produces a forecast $\hat{y} = f(q, \mathcal{I}_{t_d})$, where $f$ represents the model's forecasting function and $\mathcal{I}_{t_d}$ denotes the information set available up to deadline $t_d$.

Our dual-track taxonomy distinguishes forecasting problems primarily by \textit{temporal predictability}: recurrent forecasts target scheduled disclosures with known timing but uncertain numerical outcomes, while non-recurrent forecasts address events whose occurrence itself cannot be anticipated from calendars. Within each track, tasks are further organized by \textit{forecasting scope} into corporate and macro levels, yielding four complementary evaluation dimensions.

\subsection{Data Collection}

According to the data infrastructure provided by the Asian Institute of Digital Finance (AIDF), the Data Collection Agent continuously monitors and collects four categories of information: 1) corporate filings from regulatory databases, 2) government releases from statistical agencies, 3) financial news from real-time streams, and 4) market data from exchanges. All collected data is organized into a timestamped database and index, enabling temporal isolation during evaluation by restricting model access to content published before prediction deadlines.

\subsection{Task Generation}
\label{task}

Task generation employs two specialized agents corresponding to our dual-track taxonomy.

\noindent \textbf{Recurrent Task Generation.}
The Recurrent Task Generation Agent constructs recurrent tasks through a two-stage pipeline: identifying scheduled disclosures from official calendars, then applying template-based question generation to produce tasks with temporal parameters $(t_g, t_d, t_e)$. Macro-level tasks monitor 14 indicators (\eg GDP growth, PPI change) and corporate-level tasks target 121 financial metrics (\eg ROA, EPS). Complete specifications are provided in Appendix~\ref{app:recurrent}.

\noindent \textbf{Non-Recurrent Task Generation.}
The Non-Recurrent Task Generation Agent employs an LLM-based pipeline in three stages: (1) signal detection identifies indicators from news streams, (2) relevance assessment evaluates predictive salience, and (3) LLM-based question generation produces tasks with explicit event definitions. 
Macro tasks follow a taxonomy of 9 high-level categories and 26 fine-grained subcategories, which are instantiated via a Core–Adaptive interface into economy-specific event specifications (e.g., rate hikes, policy shifts). Corporate tasks are defined over 70 curated event types (e.g., M\&A, CEO change) with clear predictive semantics and objectively verifiable outcomes, which are instantiated at the level of individual listed companies.
Complete taxonomies are provided in Appendix~\ref{app:nonrecurrent}, and implementation details of the generation pipeline are described in Appendix~\ref{app:nonrecurrent_pipeline}.

\subsection{Forecasting}

The Forecasting Agent elicits predictions through a three-step workflow. First, \textit{task scheduling} organizes generated tasks by their prediction deadlines $t_d$ and assigns them to weekly evaluation batches. Second, \textit{model invocation} calls each evaluated model via its API with standardized prompts containing the prediction question and deadline; to ensure temporal isolation, search-augmented and deep research models are configured to access only content published before $t_d$. Third, \textit{prediction storage} records all model outputs with timestamps into a structured database for subsequent evaluation.

\subsection{Evaluation}

To ensure evaluation objectivity and simulate real-world financial accountability, we adopt a deterministic, outcome-oriented protocol managed by two agents.

\noindent\textbf{Ground Truth Extraction.}
For non-recurrent tasks, we employ a human-in-the-loop protocol to prevent evaluation bias: an LLM agent first aggregates multi-source evidence to propose potential outcomes, which are then \textbf{strictly verified by domain experts for 100\% of the samples} to determine the final ground truth. Details of this verification protocol are provided in Appendix~\ref{app:groundtruth_nonrecurrent}.

\noindent\textbf{Scoring Metric.} Each forecasting task is evaluated using a binary scoring function that awards 1 for correct forecasts and 0 otherwise:
\begin{equation}
\text{Score}(y, \hat{y}) = 
\begin{cases}
\mathbf{1}\left[\left|\frac{\hat{y} - y}{y}\right| < \epsilon_k\right] & \text{if recurrent} \\[8pt]
\mathbf{1}\left[\hat{y} = y\right] & \text{if non-recurrent}
\end{cases}
\end{equation}
where $\mathbf{1}[\cdot]$ denotes the indicator function, $y$ is the ground truth, $\hat{y}$ is the model forecast, and $\epsilon_k$ is the indicator-specific tolerance threshold. For recurrent tasks, a forecast is correct if the relative error falls within the threshold; for non-recurrent tasks, exact match is required. The overall accuracy is computed as the percentage of correct forecasts:
\begin{equation}
\text{Accuracy} = \frac{1}{N}\sum_{i=1}^{N}\text{Score}(y_i, \hat{y}_i) \times 100\%
\end{equation}
where $N$ is the total number of evaluated tasks. Results are disaggregated by task category, forecasting horizon, and market region.

\section{\bench{}}

To validate \framework, we introduce \bench{}, a weekly evaluation benchmark spanning ten weeks and targeting 8 major markets.
This section describes the construction and quality control of \bench{}, compares it with existing benchmarks, and defers detailed statistics to Appendix~\ref{sec:statistics}.

\subsection{\bench~Generation}

We instantiate \bench~using the proposed \framework~system under the following settings.

\noindent \textbf{Market and Company Selection.}
Markets are selected based on economic significance and data availability. The current instantiation spans three continents and covers \textbf{eight major economies}: US, CN, HK, JP, UK, DE, FR, and SG. To establish a standardized evaluation universe, we anchor company selection to nine leading equity indices across the covered economies (S\&P 500, NASDAQ 100, FTSE 100, DAX 40, CAC 40, Nikkei 225, CSI 300, HSI, and STI). The resulting corporate pool comprises 1,314 constituent firms, defined by index membership at a fixed snapshot date (2 Oct 2025).

\noindent \textbf{Weekly Task Generation.}
The benchmark has operated continuously since 27 October 2025, releasing a new task batch every Thursday. For recurrent corporate tasks, we employ a dynamic stratified sampling strategy, selecting up to 30\% of reporting companies per market weekly to balance density across regions; for recurrent macro tasks, we cover all scheduled indicator releases across the 8 economies. For non-recurrent tasks, candidates are generated from live news streams and undergo weekly expert review, with only the highest-quality events selected for inclusion to ensure predictive value and answerability.

\noindent \textbf{Ground Truth Acquisition.}
Ground truth acquisition operates on a rolling weekly cycle, processing tasks where the evaluation time $t_e$ has passed every Monday. For non-recurrent tasks, domain experts verify the automated classification results to ensure validity. To handle real-world irregularities such as delayed disclosures, tasks with indeterminate outcomes are marked as Pending and revisited weekly. 
Tasks remaining unresolved after a 2-week validity window are marked as \textit{Void} and excluded from performance evaluation.

\subsection{Quality Control}

We ensure benchmark reliability through systematic quality control at each stage of the pipeline.

\noindent \textbf{Expert Involvement.}
Domain experts participate throughout the benchmark lifecycle. A team of researchers with doctoral-level training in finance and economics contributes to multiple stages: designing standardized templates for recurrent task generation, defining event taxonomies and early signal criteria for non-recurrent tasks, conducting reviews of generated questions, and verifying ground truths.

\noindent \textbf{Task Quality Assurance.}
For recurrent tasks, standardized templates validated by domain experts ensure consistent metric definitions aligned with regulatory reporting standards. For non-recurrent tasks, experts review all candidate questions weekly and apply strict selection criteria, ensuring both quality and cross-market balance in the final task set.

\noindent \textbf{Ground Truth Verification.}
For recurrent tasks, ground truth is extracted from official sources through automated parsing. Sampling verification against primary sources confirms 99.8\% accuracy, with rare discrepancies attributable to subsequent data restatements by issuers. For non-recurrent tasks, ground truth determination combines automated evidence aggregation with expert validation, achieving 95\% inter-annotator agreement.

\begin{table}[t]
\caption{Comparison between \bench~and existing benchmarks. ``T-S'' and ``T-IS'' denote ``Time-Sensitive'' and ``Time-Insensitive''.}
\label{tab:benchmark_comparison}
\centering
\small 
\begin{tabularx}{\textwidth}{lXXXX} 
\toprule
 & \textbf{Domain} & \textbf{Type} & \textbf{Question} & \textbf{Frequency} \\
\midrule
\multicolumn{5}{l}{\textit{Finance Forecasting}} \\
FLUE          & Finance & T-IS & Curated        & One-time \\
PIXIU         & Finance & T-IS & Curated        & One-time \\
FinanceBench  & Finance & T-IS & Annotation     & One-time \\
FinBen        & Finance & T-IS & Curated        & One-time \\
FinCall       & Finance & T-IS & Rule-based     & One-time \\
\midrule
\multicolumn{5}{l}{\textit{Live Benchmarks}} \\
LiveBench      & Misc     & T-IS & Curated        & Monthly \\
LiveCodeBench  & Coding   & T-IS & Rule-based     & Monthly \\
SWE-bench      & Software & T-IS & Rule-based     & Seasonal \\
MathArena      & Math     & T-IS & Curated        & Seasonal \\
CryptoBench    & Crypto   & T-IS & Annotation     & Monthly \\
LiveXiv        & Scientific& T-IS & Rule-based    & Monthly \\
ForecastBench  & General  & T-S  & Rule-based     & Weekly \\
FutureX        & General  & T-S  & Rule-based     & Weekly \\
\midrule
\multicolumn{5}{l}{\textbf{Ours}} \\
\framework{} & \textbf{Finance} & \textbf{T-S} & \textbf{Dynamic} & \textbf{Weekly} \\
\bottomrule
\end{tabularx}
\end{table}

\subsection{Comparison with Existing Benchmarks}

Table~\ref{tab:benchmark_comparison} compares \bench{} with existing benchmarks across key dimensions. We make the following observations: 
1) Existing financial forecasting benchmarks primarily focus on \emph{time-insensitive} and \emph{recurrent} forecasting tasks. These are typically evaluated in a one-time setting on historical data, where the ground truth is already known at test time, leading to inherently vulnerable to data contamination.
2) Most existing live benchmarks are \emph{time-insensitive} in deterministic domains such as coding, software engineering, and mathematics, resulting in insufficiently rigorous evaluations because models may rely on the recall of pre-existing answer patterns.
3) \textit{Time-sensitive} benchmarks represent important steps toward evaluating forecasting tasks; however, they largely rely on fixed templates or rule-based extraction pipelines for task generation, limiting task diversity and constraining the benchmark’s ability to reflect evolving real-world scenarios. 

\noindent \textbf{Key Differentiators.} \bench{} differs from prior benchmarks through three characteristics:
1) a dual-track taxonomy that jointly supports recurrent and non-recurrent forecasting tasks, covering both regular disclosures and event-driven predictions;
2) financial domain specialization with objectively verifiable ground truth derived from official filings, statistical releases, and authoritative disclosures;
3) fully dynamic task generation in live market environments with weekly updates, enabling a continuous, research-oriented and contamination-free evaluation of DR agents.

\section{Experiments}
We comprehensively assess the financial forecasting capabilities of state-of-the-art methods.

\subsection{Evaluation Models}

We evaluate 13 models spanning three paradigms with distinct information access capabilities.

\headernodot{LLM with Thinking (T).}
OpenAI GPT-5 (T)~\cite{gpt5}, Claude-Sonnet-4.5 (T)~\cite{claudesonnet45}, Gemini 2.5 Pro (T)~\cite{gemini25pro}, Deepseek-v3.2 (T)~\cite{deepseekv32} and Grok 4 (T)~\cite{grok4DR}.

\headernodot{LLM with Thinking + Search (T+S).} 
OpenAI GPT-5 (T+S)~\cite{gpt5}, Claude-Sonnet-4.5 (T+S)~\cite{claudesonnet45}, Gemini 2.5 Pro (T+S)~~\cite{gemini25pro}, Deepseek-v3.2 (T+S)~~\cite{deepseekv32}, and Grok 4 (T+S)~~\cite{grok4DR}.

\headernodot{Deep Research.} OpenAI o3-deep-research~\cite{OpenAIDR}, 
Perplexity Sonar Deep Research~\cite{perplexitydeepresearch} and 
Tongyi Deep Research~\cite{tongyiDR}.

\subsection{Implementation Details}

\noindent \textbf{Temporal Isolation.}
To ensure a fair comparison, we enforce strict temporal isolation between task generation, model prediction, and performance evaluation. All evaluated models can only access the content published before the prediction deadline $t_d$, preventing access to information unavailable at prediction time.

\noindent \textbf{Task Generation.}
Tasks are generated every Thursday with prediction deadline $t_d$ set to the following Sunday 23:59 (UTC+8). For recurrent corporate tasks, we select companies with earnings releases scheduled within the prediction window. For non-recurrent tasks, domain experts review candidate questions and select about 20\% for inclusion based on prediction quality and market balance.

\noindent \textbf{Model Forecasting.} All models receive standardized prompts specifying the prediction question, deadline, and required output format. Models produce structured outputs: numerical estimates $\hat{y} \in \mathbb{R}$ for recurrent tasks and binary predictions $\hat{y} \in \{\text{Yes}, \text{No}\}$ for non-recurrent tasks. Samples of the input and output are provided in Appendix~\ref{sec:sample_forecasting}.
Model settings for forecasting tasks enable thinking and web searching capabilities where applicable. Unless otherwise specified, all parameters use default values. 
Detailed model configurations are provided in the Appendix~\ref{sec:model_settings}.

\noindent \textbf{Answer Evaluation.}
For recurrent tasks, we apply indicator-specific thresholds $\epsilon_k$ based on unit type and indicator category. By unit type, thresholds are set to 5\% for million-scale financial metrics and 1\% for percentage and ratio metrics. By indicator category, thresholds are set to 0.1\% for interest rates and foreign exchange rates, and 1\% for other macro indicators. For non-recurrent tasks, ground truth is determined through evidence aggregation from multiple sources, with expert verification achieving 95\% inter-annotator agreement.

\begin{figure}
  \centering
  \includegraphics[width=0.95\textwidth]{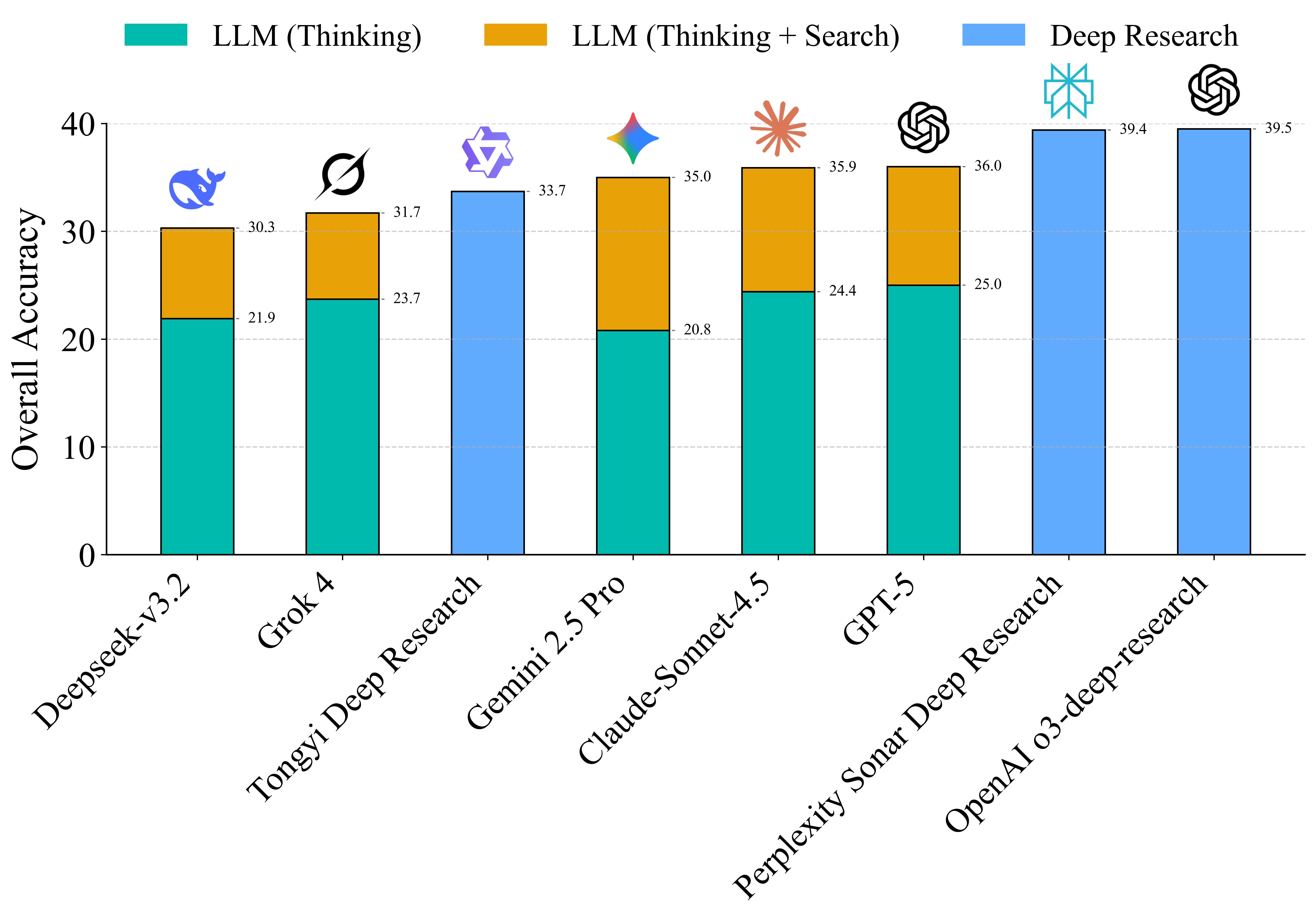}
  \caption{Main results. Overall model performance comparison over the entire ten-week horizon. }
  \label{fig:main-result}

\end{figure}

\subsection{Main Results}

We evaluate model performance across nearly 1,400 forecasting tasks generated over a 10-week period, with accuracy measured as the proportion of correct predictions. The empirical results from Figure \ref{fig:main-result} reveal three critical findings: 
1) Deep Research models establish clear performance superiority when compared to the other two counterparts. OpenAI o3-deep-research (39.5\%) and Perplexity Sonar Deep Research (39.4\%) outperform all other architectural approaches, with their near-parity suggesting convergence at the frontier of this paradigm. 
2) Augmenting reasoning with search functionality, while beneficial, cannot replicate Deep Research performance. The leading Thinking + Search (T+S) models, GPT-5 (36.0\%), Claude-Sonnet-4.5 (35.9\%), and Gemini 2.5 Pro (35.0\%), cluster within a narrow band yet consistently underperform Deep Research systems by 3-4 percentage points, indicating that search augmentation alone does not capture the full advantages of specialized Deep Research architectures. 3) External information retrieval capabilities are critical for benchmark performance. Within-model comparisons between T and T+S configurations reveal systematic accuracy degradation when search is removed, with performance drops ranging from 11.0 percentage points (GPT-5: 36.0\% → 25.0\%) to 14.2 percentage points (Gemini 2.5 Pro: 35.0\% → 20.8\%). This uniform pattern across all evaluated models provides robust evidence that access to external information is a fundamental determinant of success on this benchmark, with pure reasoning capabilities alone proving insufficient.

\begin{table}[htbp]
\centering
\caption{Performance analysis across non-recurrent (Non-rec.) and recurrent (Rec.) scenarios. Values reported are overall accuracy.}
\label{tab:tasks}
\small 
\renewcommand{\arraystretch}{1.2} 
\begin{tabular*}{\textwidth}{@{\extracolsep{\fill}} l cccccc @{}}
\toprule
 & \multicolumn{3}{c}{\textbf{Non-rec.}} & \multicolumn{3}{c}{\textbf{Rec.}} \\
\cmidrule(lr){2-4}\cmidrule(lr){5-7}
\textbf{Method} & \textbf{Corp.} & \textbf{Mac.} & \textbf{Ovr.} & \textbf{Corp.} & \textbf{Mac.} & \textbf{Ovr.} \\
\midrule

\rowcolor{lightgray!50} \multicolumn{7}{c}{\textbf{\textit{LLM (Thinking)}}} \\
OpenAI GPT-5 (T)      & 68.4 & 65.6 & 67.5 & 6.8 & 4.1 & 6.0 \\
Claude-Sonnet-4.5 (T) & 66.8 & 70.3 & 68.0 & 8.4 & 1.0 & 6.3 \\
Grok 4 (T)            & 73.7 & 71.2 & 73.1 & 11.5 & 2.7 & 9.0 \\
Deepseek-v3.2 (T)     & 61.9 & 58.6 & 60.8 & 6.9 & 2.0 & 5.6 \\
Gemini 2.5 Pro (T)    & 73.3 & 68.8 & 71.7 & 8.5 & 1.0 & 6.4 \\

\rowcolor{lightgray!50} \multicolumn{7}{c}{\textbf{\textit{LLM (Thinking + Search)}}} \\
OpenAI GPT-5 (T+S)      & 78.1 & 72.7 & 76.3 & 22.8 & 11.1 & 19.5 \\
Claude-Sonnet-4.5 (T+S) & 79.8 & 73.4 & 77.6 & 20.7 & 19.6 & 20.4 \\
Grok 4 (T+S)            & 74.5 & \underline{77.3} & 75.5 & 15.1 & 18.2 & 16.0 \\
Deepseek-v3.2 (T+S)     & 76.9 & 70.3 & 74.7 & 13.4 & 14.6 & 13.7 \\
Gemini 2.5 Pro (T+S)    & 78.5 & \bf 77.5 & 78.3 & 23.3 & 17.6 & 21.7 \\

\rowcolor{lightgray!50} \multicolumn{7}{c}{\textbf{\textit{Deep Research}}} \\
Perplexity Sonar      & \bf 81.4 & 75.0 & \bf 79.2 & \underline{26.2} & \bf 23.7 & \bf 25.5 \\
Tongyi Deep Research  & 79.8 & 74.2 & 77.9 & 23.5 & 15.5 & 21.2 \\
OpenAI o3-deep        & \underline{80.6} & 75.8 & \underline{78.9} & \bf 26.7 & \underline{21.3} & \underline{25.2} \\

\bottomrule
\end{tabular*}
\end{table}

\begin{table}[htbp]
\centering
\caption{The performance analysis across 8 financial markets. The values reported denote the overall accuracy.}
\label{tab:markets}
\small
\renewcommand{\arraystretch}{1.2}
\begin{tabular*}{\textwidth}{@{\extracolsep{\fill}} l cccccccc @{}}
\toprule
\multirow{2}{*}{\textbf{Method}} 
 & \includegraphics[width=0.35cm]{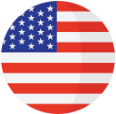} & \includegraphics[width=0.35cm]{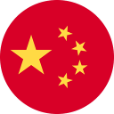} & \includegraphics[width=0.35cm]{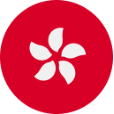} & \includegraphics[width=0.35cm]{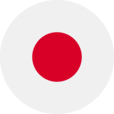} & \includegraphics[width=0.35cm]{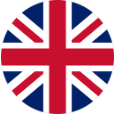} & \includegraphics[width=0.35cm]{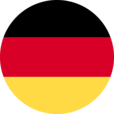} & \includegraphics[width=0.35cm]{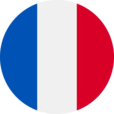} & \includegraphics[width=0.35cm]{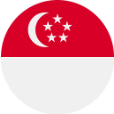} \\
 & \textbf{US} & \textbf{CN} & \textbf{HK} & \textbf{JP} & \textbf{UK} & \textbf{DE} & \textbf{FR} & \textbf{SG} \\
\midrule

\rowcolor{lightgray!50} \multicolumn{9}{c}{\textbf{\textit{LLM (Thinking)}}} \\
OpenAI GPT-5 (T)      & 19.0 & 32.3 & 26.7 & 14.4 & 28.6 & 47.6 & 39.1 & 32.6 \\
Claude-Sonnet-4.5 (T) & 18.1 & 28.5 & 28.3 & 16.7 & 31.0 & 44.0 & 32.7 & 32.6 \\
Grok 4 (T)            & 20.2 & 26.9 & 20.8 & 17.5 & 31.0 & 39.3 & 26.4 & 31.5 \\
Deepseek-v3.2 (T)     & 16.5 & 26.9 & 25.8 & 12.5 & 27.8 & 42.9 & 30.9 & 28.1 \\
Gemini 2.5 Pro (T)    & 18.7 & 24.6 & 19.2 & 12.8 & 26.2 & 39.3 & 21.8 & 25.8 \\

\rowcolor{lightgray!50} \multicolumn{9}{c}{\textbf{\textit{LLM (Thinking + Search)}}} \\
OpenAI GPT-5 (T+S)      & 36.7 & 40.0 & 35.8 & 22.6 & 38.9 & 47.6 & \underline{44.5} & 39.3 \\
Claude-Sonnet-4.5 (T+S) & 33.4 & 42.3 & \underline{42.5} & 23.7 & \underline{39.7} & 48.8 & 40.0 & 43.8 \\
Grok 4 (T+S)            & 29.2 & 40.8 & 34.2 & 16.7 & 37.3 & 45.2 & 37.3 & \underline{44.9} \\
Deepseek-v3.2 (T+S)     & 26.4 & 36.9 & 40.3 & 17.1 & 33.3 & 45.8 & 37.3 & 39.3 \\
Gemini 2.5 Pro (T+S)    & 35.7 & 38.0 & 30.8 & 22.0 & 34.7 & 47.5 & 44.5 & \bf 47.7 \\

\rowcolor{lightgray!50} \multicolumn{9}{c}{\textbf{\textit{Deep Research}}} \\
Perplexity Sonar      & \underline{40.1} & \bf 45.7 & \bf 47.9 & \underline{24.3} & \bf 40.0 & \underline{50.0} & 40.9 & \underline{44.9} \\
Tongyi Deep Research  & 33.9 & 44.6 & 29.4 & 20.7 & 38.7 & 47.0 & 36.4 & 36.4 \\
OpenAI o3-deep        & \bf 41.0 & \underline{45.4} & 40.0 & \bf 25.3 & 38.9 & \bf 51.2 & \bf 48.2 & 41.6 \\

\bottomrule
\end{tabular*}
\end{table}

\subsection{In-depth Analysis}

\headernodot{Performance Analysis on Different Tasks.}
\label{sec:diff_tasks}
We analyze the model performance across different scenarios (\ie recurrent and non-recurrent) at different levels (\ie corporate and macro).
We present the results in Table \ref{tab:tasks}, from which we make the following key findings: 
1) The thinking-only LLMs achieve reasonable accuracy on non-recurrent tasks but collapse on recurrent scenarios (often below 10\% overall), suggesting that internal reasoning alone is insufficient for temporally grounded, fine-grained financial prediction.
2) LLMs augmented with thinking and search capabilities outperform thinking-only counterparts across all task types, indicating the importance of external information access. However, their improvements on recurrent tasks remain modest, implying that information retrieval alone is insufficient for addressing the challenges in \bench.
4) Deep Research agents achieve the best performance across both non-recurrent and recurrent tasks, particularly on recurrent corporate and macro forecasting. This suggests that multi-step planning, evidence synthesis, and structured reasoning jointly contribute to stronger forecasting under strict temporal isolation.
4) Across all paradigms, models achieve high accuracy on non-recurrent tasks (up to 81.4\%), while performance on recurrent tasks drops sharply, with the best method reaching only 25.5\% overall. This highlights the intrinsic difficulty of precise, numeric forecasting under periodic disclosure compared to binary event prediction.
To better understand the failure modes, we provide a error case study in Appendix~\ref{sec:error_case_study}.

\headernodot{Performance Analysis across Different Markets.}
Then, we analyze the model performance across different markets, and present the results in Table \ref{tab:markets}, from which we observe:
1) Deep Research agents achieve the highest accuracy in nearly all markets, indicating superior generalization across heterogeneous regulatory regimes, disclosure standards, and information environments.
2) In every market, adding search capabilities leads to substantial gains, underscoring the importance of accessing up-to-date and market-specific information in live financial forecasting settings.
3) Most methods perform best in information-rich markets such as the US and China, while accuracy is consistently lower in markets with relatively less available data or language diversity, such as Japan.

\begin{figure}[t] 
  \centering
  \setlength{\abovecaptionskip}{-0.3pt}
  \setlength{\belowcaptionskip}{-0.3cm}
  \includegraphics[width=0.95\textwidth]{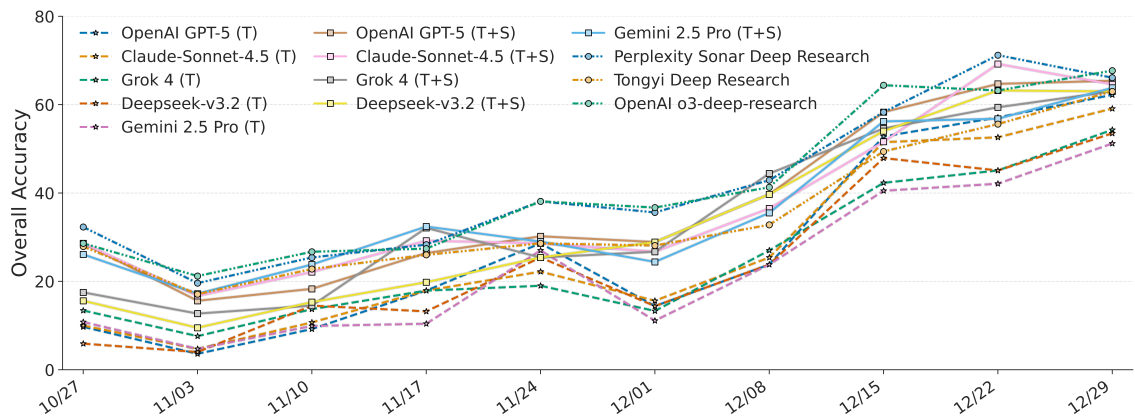}
  \caption{Weekly performance comparison.}
  \label{fig:weeks}
\end{figure}

\headernodot{Weekly Performance Analysis.}
We analyze the model performance over all ten weeks, and present the results in Figure \ref{fig:weeks}. We can observe: 
1) Accuracy improves steadily across weeks as the proportion of recurrent tasks declines following the end of the disclosure period, consistent with the stronger performance of all models on non-recurrent tasks shown in Table \ref{tab:tasks}.
2) Deep Research agents outperform all other methods consistently across all weeks, indicating a superior and stable capacity to integrate observed signals and adapt over time.

\section{Related Work}
\subsection{Deep Research Agents}
Deep Research (DR) agents aim to solve complex tasks through planning, information gathering, and multi-step reasoning, and have recently been widely deployed in both industrial and open-source LLM systems \cite{openai2025deepresearch, citron2025deepresearch, qwen2025deepresearch, perplexity2025deepresearch, team2025tongyi}. Early works \cite{yao2023react, shinn2023reflexion} first established core agentic paradigms that interleave reasoning with tool use and environment interaction. This was later extended to realistic domains, such as web-based information seeking and software engineering \cite{zhou2023webarena, huang2023agentcoder}. In parallel, benchmarks \cite{liu2023agentbench,zhu2025findeepresearch} were also proposed to evaluate agentic capabilities in controlled environments. However, most existing evaluations rely on static task sets, limiting their ability to capture agent behavior under changing environments. As DR agents increasingly operate in evolving real-world contexts, reliably evaluating performance under dynamic and time-sensitive conditions has become a critical challenge that our work seeks to address.

\subsection{Live Benchmarks}
Live benchmarking for LLMs has emerged as a key direction for mitigating data contamination in evaluation \cite{balloccu2024leak}. Existing live benchmarks can be broadly categorized into \textit{time-insensitive} and \textit{time-sensitive} tasks. While time-insensitive benchmarks \cite{white2025livebench, jain2024livecodebench, zhang2025swe, balunovic2025matharena} seek to mitigate contamination through continuous updates, their tasks rely on deterministic ground truths that do not depend on future outcomes. In contrast, \textit{time-sensitive} benchmarks \cite{karger2024forecastbench, zeng2025futurex} evaluate predictive reasoning on problems whose answers are unknown at test time. However, these benchmarks typically rely on manual curation or rule-based extraction pipelines for task construction, which constrains task diversity and adaptability. Our work introduces a time-sensitive benchmark for financial forecasting, where tasks are dynamically generated from evolving real-world market environments.

\subsection{Financial Forecasting Benchmarks}
\vspace{-1px}
Current financial forecasting benchmarks \cite{shah2022flue, islam2023financebench} typically focus on \textit{recurrent} events, which are regularly occurring targets such as stock price movements \cite{xie2023pixiu, xie2024finben} or company earnings \cite{shu2025fincall}. These benchmarks are constructed from static historical datasets, which trigger concerns over data contamination \cite{sainz2023nlp} from LLM-based solutions. Another crucial but less common benchmark is on \textit{non-recurrent} tasks, which are discrete events that are also known to impact financial markets \cite{hu2018listening, koa2024learning}, such as new partnerships or tariffs. The benchmarks can also be categorized into \textit{corporate-level} \cite{zhu2025fincast} or \textit{macro-level} \cite{zhang2025camef} tasks, which differ in terms of event scale and granularity. Our work deals with these tasks under a unified evaluation framework, constructed from live data. 

\section{Conclusion}

In this work, we introduce \framework{}, the first live, end-to-end multi-agent system for evaluating DR agents in financial forecasting. 
It can continuously generate forward-looking, research-oriented tasks under strict temporal isolation, and integrates task creation, model invocation, and ground-truth verification into a unified and fully automated pipeline.
With this system, we instantiate \bench{}, a weekly benchmark covering recurrent numerical disclosures and non-recurrent event-driven predictions at both corporate and macroeconomic levels. 
We evaluate 13 representative systems, demonstrating that current DR agents are still challenged by genuinely research-oriented financial forecasting, particularly in precise recurrent numerical forecasting.
\framework{} establishes a dynamic and contamination-free evaluation paradigm for DR agents in live market environments and provides a foundation on which future systems and benchmarks can be continuously built and extended.

In future, we plan to extend the system to richer task forms, including probabilistic, multi-step, and portfolio-level forecasting, and to incorporate process-based evaluation to better understand how DR agents search, reason, and fail in live forecasting scenarios.

\section{Contributions}

\begin{itemize}
    \item \textbf{Project Leader:} Fengbin Zhu
    \item  \textbf{Major Contributors:} Xiangyu Li, Xuan Yao, Guohao Qi
    \item  \textbf{Secondary Contributors:} Kelvin J.L. Koa, Xiang Yao Ng, Ziyang Liu, Xingyu Ni, Chang Liu, Yonghui Yang, Yang Zhang 
    \item  \textbf{Advisors:} Wenjie Wang, Fuli Feng, Chao Wang, Huanbo Luan, Xiaofen Xing, Xiangmin Xu,   Tat-Seng Chua, and Ke-Wei Huang.
\end{itemize}

\clearpage


\bibliographystyle{ACM-Reference-Format}
\bibliography{arxiv/FinDeepForecast}

\clearpage

\appendix
\label{sec:appendix}

\section{Recurrent Task Specifications}
\label{app:recurrent}

\subsection{Macro Indicators}

We monitor 96 macro indicators, derived from 14 indicator types that are instantiated across eight economies and complemented by two global market indices, as detailed in Table~\ref{tab:macro_indicators}.
The selection criteria capture the four fundamental pillars of macro analysis: real economic activity (e.g., GDP, Unemployment), price stability (e.g., CPI, PPI), monetary conditions (e.g., Interest Rates, Stock Index), and external balance (e.g., FX Rate, CAB). Crucially, we augment these with global barometers—specifically Commodities and VIX—to test the model's sensitivity to cross-border supply shocks and systemic risk sentiment. Accurately forecasting these indicators requires financial experts to conduct extensive information gathering and multi-step reasoning, making them ideal proxies for evaluating deep research capabilities.

\begin{table}[h]
\centering
\caption{Macro indicators for recurrent tasks.}
\label{tab:macro_indicators}
\scriptsize
\begin{tabularx}{\linewidth}{c|l|X|l} 
\toprule
\textbf{No.} & \textbf{Indicator} & \textbf{Description} & \textbf{Economies} \\
\midrule
\multicolumn{4}{c}{\cellcolor{lightgray}\textbf{Global Indices (2)}} \\
\midrule
1 & S\&P GSCI Commodity & Global commodity price index & Global \\
2 & CBOE VIX & Market volatility index & Global \\
\midrule
\multicolumn{4}{c}{\cellcolor{lightgray}\textbf{Economy-Specific Indicators (94)}} \\
\midrule
3 & Stock Index & Major equity market index & All 8 \\
4 & Interest Rate (1yr) & 1-year govt bond yield & Excl. SG \\
5 & Interest Rate (3m) & 3-month treasury bill rate & All 8 \\
6 & FX Rate & Exchange rate against USD & Excl. US \\
7 & GDP & Gross Domestic Product & All 8 \\
8 & CPI & Consumer Price Index & All 8 \\
9 & PPI & Producer Price Index & All 8 \\
10 & UNRATE & Unemployment Rate & All 8 \\
11 & HPI & House Price Index & All 8 \\
12 & NEER & Nominal Effective Exchange Rate & All 8 \\
13 & Interbank Rate & 3-month interbank lending rate & All 8 \\
14 & CAB & Current Account Balance & All 8 \\
\midrule
\multicolumn{4}{l}{\textbf{Economies:} US, CN, HK, JP, UK, DE, FR, SG} \\
\multicolumn{4}{l}{\textbf{Note:} US excludes FX Rate; SG excludes Interest Rate (1yr)} \\
\bottomrule
\end{tabularx}
\end{table}

\subsection{Corporate Financial Metrics}

We cover 121 corporate financial metrics organized into 9 categories, as shown in Table~\ref{tab:corporate_metrics_1} and Table~\ref{tab:corporate_metrics_2}. This extensive selection mirrors the comprehensive framework used in professional fundamental analysis. By encompassing the three primary financial statements (Balance Sheet, Income Statement, Cash Flow) and six categories of derived ratios, we require the model to not only retrieve raw data but also perform arithmetic reasoning to assess liquidity, solvency, and operational efficiency. Accurately forecasting these metrics typically requires financial experts to conduct detailed information gathering and multi-step analytical reasoning, providing a rigorous test of deep research capabilities.

\begin{table*}[h]
\centering
\caption{Corporate financial metrics - Part 1 (Balance Sheet, Income Statement, Cash Flow).}
\label{tab:corporate_metrics_1}
\small
\setlength{\tabcolsep}{0.9mm}
\resizebox{\textwidth}{!}{
\begin{tabular}{c|l|l||c|l|l}
\toprule
\textbf{No.} & \textbf{Metric} & \textbf{Description} & \textbf{No.} & \textbf{Metric} & \textbf{Description} \\
\midrule
\multicolumn{6}{c}{\cellcolor{lightgray}\textbf{Balance Sheet Items (25 metrics)}} \\
\midrule
1 & Total Assets & Total value of all assets & 14 & Accounts Payable & Amounts owed to suppliers \\
2 & Total Liabilities & Total liabilities owed & 15 & Accrued Expenses & Expenses incurred but not paid \\
3 & Total Equity & Total shareholders' equity & 16 & Deferred Revenue & Revenue received but not earned \\
4 & Total Current Assets & Assets convertible within 1 yr & 17 & Retained Earnings & Accumulated undistributed income \\
5 & Total Current Liabilities & Liabilities due within 1 yr & 18 & Treasury Stock & Repurchased company shares \\
6 & Long Term Debt & Debt due beyond 1 year & 19 & Minority Interest & Non-controlling interest \\
7 & Short Term Debt & Debt due within 1 year & 20 & Preferred Stock & Preferential dividend equity \\
8 & Short and Long Term Debt & Combined total debt & 21 & Common Stock & Basic ownership shares \\
9 & Total Loans & Loans held by financials & 22 & Total Deposits & Customer deposits (financials) \\
10 & Cash and Equivalents & Liquid assets and cash & 23 & Saving Deposits & Savings account deposits \\
11 & Accounts Receivable & Amounts owed by customers & 24 & Property Plant \& Equip. & Physical asset value \\
12 & Inventory & Goods held for sale & 25 & Intangible Assets & Patents, goodwill, etc. \\
13 & Goodwill & Acquisition premium paid & & & \\
\midrule
\multicolumn{6}{c}{\cellcolor{lightgray}\textbf{Income Statement Items (22 metrics)}} \\
\midrule
26 & Revenue & Total sales and income & 37 & Interest Income & Income from interest \\
27 & Cost of Revenue & Direct costs of goods sold & 38 & Other Income & Non-operating income \\
28 & Gross Profit & Revenue minus cost of revenue & 39 & Extraordinary Items & Unusual gains or losses \\
29 & Operating Income & Profit from core operations & 40 & Discontinued Operations & Results from closed segments \\
30 & EBIT & Earnings before interest \& taxes & 41 & EPS (Basic) & Net income per basic share \\
31 & EBITDA & EBIT plus depreciation \& amort. & 42 & EPS (Diluted) & Net income per diluted share \\
32 & Net Income & Final profit after all expenses & 43 & Dividends Per Share & Dividends paid per share \\
33 & Interest Expense & Interest paid on debt & 44 & Revenue Growth (YoY) & Year-over-year revenue change \\
34 & R\&D Expense & Research \& development costs & 45 & Net Income Growth (YoY) & Year-over-year income change \\
35 & SG\&A Expense & Selling, general \& admin costs & 46 & Operating Expense & Total operating costs \\
36 & Income Tax Expense & Taxes on corporate income & 47 & Pre-Tax Income & Income before tax expense \\
\midrule
\multicolumn{6}{c}{\cellcolor{lightgray}\textbf{Cash Flow Items (15 metrics)}} \\
\midrule
48 & Cash From Operations & Net cash from operating & 56 & Debt Repayment & Cash used to repay debt \\
49 & Cash from Investing & Net cash from investing & 57 & Debt Issuance & Cash from issuing debt \\
50 & Cash from Financing & Net cash from financing & 58 & Stock Repurchase & Cash for share buybacks \\
51 & Free Cash Flow & Operating cash minus capex & 59 & Stock Issuance & Cash from issuing shares \\
52 & Depreciation \& Amort. & Non-cash asset reduction & 60 & Dividend Payments & Cash paid as dividends \\
53 & Capital Expenditure & Investment in fixed assets & 61 & Net Change in Cash & Total cash position change \\
54 & Acquisitions & Cash for acquiring companies & 62 & Working Capital Changes & Op. asset/liability changes \\
55 & Divestitures & Cash from selling units & & & \\
\bottomrule
\end{tabular}
}
\end{table*}

\begin{table*}[h]
\centering
\caption{Corporate financial metrics - Part 2 (Profitability, Liquidity, Leverage, Efficiency, Coverage, Valuation).}
\label{tab:corporate_metrics_2}
\small
\setlength{\tabcolsep}{0.9mm}
\resizebox{\textwidth}{!}{
\begin{tabular}{c|l|l||c|l|l}
\toprule
\textbf{No.} & \textbf{Metric} & \textbf{Description} & \textbf{No.} & \textbf{Metric} & \textbf{Description} \\
\midrule
\multicolumn{6}{c}{\cellcolor{lightgray}\textbf{Profitability Ratios (15 metrics)}} \\
\midrule
63 & Return on Assets (ROA) & Net Income / Total Assets & 71 & Return on Sales & Op. Income / Revenue \\
64 & Return on Equity (ROE) & Net Income / Total Equity & 72 & Cash Return on Assets & Op. Cash Flow / Total Assets \\
65 & Return on Invested Capital & NOPAT / Invested Capital & 73 & Cash Return on Equity & Op. Cash Flow / Total Equity \\
66 & Gross Margin & Gross Profit / Revenue & 74 & NPL Ratio & Non-Performing Loans / Loans \\
67 & Operating Margin & Op. Income / Revenue & 75 & Net Interest Margin & Net Int. Inc. / Earning Assets \\
68 & EBITDA Margin & EBITDA / Revenue & 76 & Efficiency Ratio & Non-Int. Exp. / Revenue \\
69 & Net Margin & Net Income / Revenue & 77 & Cost-to-Income Ratio & Op. Exp. / Op. Income \\
70 & Profit Margin & (Op. Inc. $-$ D\&A) / Rev. & & & \\
\midrule
\multicolumn{6}{c}{\cellcolor{lightgray}\textbf{Liquidity Ratios (8 metrics)}} \\
\midrule
78 & Current Ratio & Curr. Assets / Curr. Liab. & 82 & Working Capital & Curr. Assets $-$ Curr. Liab. \\
79 & Quick Ratio & (Curr. Assets $-$ Inv.) / Curr. Liab. & 83 & Working Capital Ratio & Working Capital / Total Assets \\
80 & Cash Ratio & Cash \& Equiv. / Curr. Liab. & 84 & Defensive Interval Ratio & Liquid Assets / Daily Op. Exp. \\
81 & Op. Cash Flow Ratio & Op. Cash Flow / Curr. Liab. & 85 & Cash Conversion Cycle & DIO + DSO $-$ DPO \\
\midrule
\multicolumn{6}{c}{\cellcolor{lightgray}\textbf{Leverage Ratios (12 metrics)}} \\
\midrule
86 & Debt-to-Equity Ratio & Total Debt / Total Equity & 92 & Long-term Debt to Assets & LT Debt / Total Assets \\
87 & Debt-to-Assets Ratio & Total Debt / Total Assets & 93 & ST Debt to Total Debt & ST Debt / Total Debt \\
88 & Liability-to-Assets Ratio & Total Liab. / Total Assets & 94 & Net Debt & Total Debt $-$ Cash \& Equiv. \\
89 & Equity Ratio & Total Equity / Total Assets & 95 & Net Debt to Equity & Net Debt / Total Equity \\
90 & Equity Multiplier & Total Assets / Total Equity & 96 & Net Debt to EBITDA & Net Debt / EBITDA \\
91 & Long-term Debt to Equity & LT Debt / Total Equity & 97 & Financial Leverage & Avg. Assets / Avg. Equity \\
\midrule
\multicolumn{6}{c}{\cellcolor{lightgray}\textbf{Efficiency Ratios (12 metrics)}} \\
\midrule
98 & Asset Turnover & Revenue / Total Assets & 104 & Equity Turnover & Revenue / Total Equity \\
99 & Fixed Asset Turnover & Revenue / Fixed Assets & 105 & Days Inventory Outstanding & 365 / Inventory Turnover \\
100 & Inventory Turnover & COGS / Avg. Inventory & 106 & Days Sales Outstanding & 365 / Receivables Turnover \\
101 & Receivables Turnover & Revenue / Avg. Receivables & 107 & Days Payables Outstanding & 365 / Payables Turnover \\
102 & Payables Turnover & COGS / Avg. Payables & 108 & Total Loans Growth (YoY) & YoY change in total loans \\
103 & Working Capital Turnover & Revenue / Working Capital & 109 & Deposits Growth (YoY) & YoY change in deposits \\
\midrule
\multicolumn{6}{c}{\cellcolor{lightgray}\textbf{Coverage Ratios (6 metrics)}} \\
\midrule
110 & Interest Coverage (EBIT) & EBIT / Interest Expense & 113 & Interest Coverage (Op. Inc.) & Op. Inc. / Interest Exp. \\
111 & Interest Coverage (EBITDA) & EBITDA / Interest Expense & 114 & Debt Service Coverage & Op. Inc. / Debt Service \\
112 & Interest Coverage (Net Inc.) & Net Income / Interest Exp. & 115 & Fixed Charge Coverage & (EBIT+Lease) / (Int.+Lease) \\
\midrule
\multicolumn{6}{c}{\cellcolor{lightgray}\textbf{Valuation \& Market Metrics (6 metrics)}} \\
\midrule
116 & Book Value Per Share & Total Equity / Shares Out. & 119 & Cash Flow Per Share & Op. Cash Flow / Shares Out. \\
117 & Tangible Book Value/Share & Tangible Equity / Shares Out. & 120 & Enterprise Value & Market Cap + Debt $-$ Cash \\
118 & Revenue Per Share & Revenue / Shares Out. & 121 & Market Capitalization & Price $\times$ Shares Outstanding \\
\bottomrule
\end{tabular}
}
\end{table*}

\section{Non-Recurrent Task Specifications}
\label{app:nonrecurrent}

\subsection{Non-Recurrent Macro Event Taxonomy}
\label{app:taxonomy_details}

To categorize non-recurrent macroeconomic shocks, we adopted a \textbf{"Stable Core, Adaptive Interface"} design philosophy. This hierarchical framework consists of a fixed semantic layer (Level 1) ensuring consistency, and a dynamic grounding layer (Level 2) ensuring relevance.

\noindent \textbf{Level 1: The Stable Semantic Taxonomy (Immutable Layer).}
The first level defines a standardized ontology of macro-financial events designed to remain invariant across time and regions. We categorize events into 9 categories (A to I) and 26 subcategories (Table~\ref{tab:full_taxonomy}). This taxonomy follows the \textit{Mutually Exclusive and Collectively Exhaustive (MECE)} principle. By keeping this semantic layer static, we ensure that model performance remains comparable across different eras, providing a consistent benchmark for longitudinal evaluation.

\noindent \textbf{Level 2: The Economy-Specific Grounding (Adaptive Layer).}
Unlike the static Level 1, the grounding layer is designed to be dynamic and extensible. This layer maps the universal concepts to specific, falsifiable market indicators for each economy.

\noindent Our expert panel designed this layer with two degrees of flexibility to accommodate the evolving nature of financial markets:
\begin{enumerate}
    \item \textbf{Dynamic Calibration:} The quantifiable triggers (e.g., specific basis point thresholds) are subject to periodic recalibration. As market regimes shift (e.g., from a low-interest environment to a high-inflation era), these parameters can be updated to maintain their discriminatory power without altering the Level 1 definitions.
    \item \textbf{Extensibility for Future Tasks:} The framework supports the seamless integration of new economies or additional event types. Future iterations of the benchmark can introduce new task subcategories or expand to emerging markets by simply defining the corresponding Level 2 grounding logic, preserving the integrity of the overarching taxonomy.
\end{enumerate}

\noindent For the current version, we defined the "Ground Truth" for 8 major economies. For each economy-subcategory pair, we established  \textbf{Authoritative Sources} which strictly designated official sources (e.g., FOMC, PBoC, OBR), and \textbf{Quantifiable Triggers} with rigid quantitative thresholds (e.g., $\ge 25$ bps rate hike, $>1\%$ GDP fiscal impulse) tailored to local market structures (e.g., "Shunto" for Japan, "Schuldenbremse" for Germany). The detailed grounding tables are presented in Tables~\ref{tab:grounding_us} through \ref{tab:grounding_sg}. This scientific design ensures that the benchmark remains a "living" evaluation standard, capable of evolving alongside the real-world financial landscape.

\small
\setlength{\tabcolsep}{1.2mm}
\renewcommand{\arraystretch}{1.35}

\small
\setlength{\tabcolsep}{1.2mm}
\renewcommand{\arraystretch}{1.35}

\begin{xltabular}{\linewidth}{p{2.2cm}|X||p{2.2cm}|X}

\caption{Taxonomy of Non-Recurrent Macro Events.} \label{tab:full_taxonomy} \\
\toprule
\textbf{Category} & \textbf{Code \& Detailed Description} & \textbf{Category} & \textbf{Code \& Detailed Description} \\
\midrule
\endfirsthead

\multicolumn{4}{c}{{\bfseries \tablename\ \thetable{} -- continued from previous page}} \\
\toprule
\textbf{Category} & \textbf{Code \& Detailed Description} & \textbf{Category} & \textbf{Code \& Detailed Description} \\
\midrule
\endhead

\midrule
\multicolumn{4}{r}{{Continued on next page...}} \\
\endfoot

\bottomrule
\endlastfoot

\multirow{5}{*}{\shortstack[l]{\textbf{A. Monetary \&}\\ \textbf{Financial} \\ \textbf{Conditions}\\(3 types)}}
& \textbf{A1 Monetary Policy Shift}: Central bank policy rate hikes/cuts, policy stance changes, quantitative easing or tapering decisions.
& \multirow{6}{*}{\shortstack[l]{\textbf{E. Real} \\ \textbf{Economy}\\ \textbf{Activity}\\(4 types)}}
& \textbf{E1 Industrial Production / Manufacturing Shock}: Shocks to industrial production or manufacturing activity, including sharp contractions or surges. \\

& \textbf{A2 Financial Market Liquidity Shock}: Bond- or money-market stress, funding squeezes, repo-market dislocations, impaired market-making.
& & \textbf{E2 Retail / Consumption / Services Shock}: Shocks to household consumption, retail sales, or services-sector activity driven by income or sentiment changes. \\

& \textbf{A3 Macro-prudential Regulation Change}: Changes to macro-prudential tools such as LTV/DTI limits, countercyclical capital buffers, or leverage caps.
& & \textbf{E3 Housing / Real Estate Cycle Shock}: Downturns or booms in property markets, construction activity, or related policy changes. \\

\cmidrule{1-2}
\multirow{5}{*}{\shortstack[l]{\textbf{B. Fiscal Policy} \\ \textbf{\&}\\ \textbf{Public Finance}\\(2 types)}}
& \textbf{B1 Fiscal Stimulus / Austerity}: Government budget decisions that significantly expand or contract spending, transfer programs, or tax burdens.
& & \textbf{E4 Technology, Digital Economy \& AI-Driven Industrial Activity}: Real-economy impacts arising from major technology/AI developments, adoption waves, or semiconductor constraints. \\

\cmidrule{3-4}
& \textbf{B2 Sovereign Debt Stress}: Events indicating sovereign credit stress, including rating downgrades, refinancing pressure, or default risk.
& \multirow{5}{*}{\shortstack[l]{\textbf{F. Financial} \\  \textbf{Stability}\\ \textbf{\& Credit} \\ \textbf{Cycle}\\(3 types)}}
& \textbf{F1 Credit Cycle Shift (Boom/Bust)}: Rapid expansions or contractions in private credit to households or corporates. \\

\cmidrule{1-2}
\multirow{6}{*}{\shortstack[l]{\textbf{C. Trade \&}\\ \textbf{External Sector}\\(3 types)}}
& \textbf{C1 Trade Policy Change / Sanctions / Tariffs}: Introduction or removal of tariffs, quotas, export controls, sanctions, or anti-dumping/countervailing duties.
& & \textbf{F2 Banking System Stress / NPL Shock}: Deterioration in bank asset quality, rising non-performing loans, or liquidity/solvency concerns. \\

& \textbf{C2 Currency / FX Pressure Shock}: Sharp exchange-rate moves, reserve losses, or capital outflows indicating FX market pressure.
& & \textbf{F3 Asset Price Shock (Equity/Bond/Volatility)}: Sharp corrections in equity or bond markets, volatility spikes, or broad market repricing events. \\

\cmidrule{3-4}
& \textbf{C3 External Financing / Current-Account Shock}: Stress related to external financing, current-account imbalances, sudden stops, or debt rollover risks.
& \multirow{5}{*}{\shortstack[l]{\textbf{G. Structural} \\ \textbf{\&}\\ \textbf{Regulatory} \\ \textbf{Policy}\\(3 types)}}
& \textbf{G1 Climate / Carbon / ESG Policy}: Policy changes related to climate targets, carbon pricing, emissions trading, or ESG disclosure rules. \\

\cmidrule{1-2}
\multirow{5}{*}{\shortstack[l]{\textbf{D. Commodity, }\\ \textbf{Energy}\\ \textbf{\& Supply} \\ \textbf{Chain}\\(3 types)}}
& \textbf{D1 Energy Price Shock}: Large and rapid changes in energy prices (oil, gas, electricity) affecting production costs and inflation.
& & \textbf{G2 Tech/Data/Privacy Regulation}: New or revised regulations on data privacy, cybersecurity, cross-border data flows, or digital governance. \\

& \textbf{D2 Commodity Price Shock}: Significant volatility in key non-energy commodities such as metals, food, or agricultural inputs.
& & \textbf{G3 Structural / Institutional Reform}: Reforms to labour markets, pensions, social security, legal or institutional frameworks. \\

\cmidrule{3-4}
& \textbf{D3 Global Supply Chain Disruption}: Logistics bottlenecks, shipping disruptions, or trade chokepoints that impair global supply chains.
& \multirow{5}{*}{\shortstack[l]{\textbf{I. Geopolitical} \\ \textbf{\&}\\ \textbf{Systemic} \\ \textbf{Shocks}\\(3 types)}}
& \textbf{I1 Conflict / Sanctions Shock}: Military conflicts, geopolitical escalation, or sanctions regimes with macro/sectoral impact. \\

\cmidrule{1-2}
\multirow{5}{*}{\shortstack[l]{\textbf{H. Labour } \\ \textbf{Market \&}\\ \textbf{Household} \\ \textbf{Sector}\\(2 types)}}
& \textbf{H1 Labour Market Shock}: Shocks to employment, unemployment, labour-force participation, or wage dynamics.
& & \textbf{I2 Natural Disaster / Pandemic Shock}: Major natural disasters or health crises that disrupt economic activity. \\

& \textbf{H2 Household Income / Consumption Stress}: Stress in household balance sheets, including real income declines, debt distress, or demand weakness.
& & \textbf{I3 Global Financial Contagion}: Spillovers from global financial crises, cross-border banking stress, or systemic liquidity shocks. \\

\end{xltabular}

\begin{xltabular}{\linewidth}{l|p{3.5cm}|X}
\caption{Event grounding standards for the United States (US) market.} \label{tab:grounding_us} \\

\toprule
\textbf{Code} & \textbf{Authority / Source} & \textbf{Quantifiable Trigger / Definition} \\
\midrule
\endfirsthead

\multicolumn{3}{c}%
{{\bfseries \tablename\ \thetable{} -- continued from previous page}} \\
\toprule
\textbf{Code} & \textbf{Authority / Source} & \textbf{Quantifiable Trigger / Definition} \\
\midrule
\endhead

\midrule
\multicolumn{3}{r}{{Continued on next page...}} \\
\endfoot

\bottomrule
\endlastfoot

\multicolumn{3}{c}{\cellcolor[gray]{0.95}\textbf{A. Monetary \& Financial Conditions}} \\
\midrule
A1 & Federal Reserve (FOMC) & Federal Funds Target Range upper bound changes by $\ge 25$ bps; OR official statement explicitly pivots stance. \\
A2 & Fed H.4.1 / NY Fed & FRA-OIS spread $>95$th percentile; OR Reverse Repo Facility usage surges $>\$500$B in a week. \\
A3 & Fed Board / FDIC & Implementation of new capital rules (e.g., Basel III Endgame) or change in CCAR stress test scenarios. \\
\midrule

\multicolumn{3}{c}{\cellcolor[gray]{0.95}\textbf{B. Fiscal Policy \& Public Finance}} \\
\midrule
B1 & CBO / White House & Passage of legislation (e.g., CARES Act, IRA) with discretionary spending impact $\ge 1\%$ of GDP. \\
B2 & Treasury / CDS Market & US Sovereign CDS (5Y) spread $>50$ bps; OR 'Extraordinary Measures' exhausted date approaches within 30 days. \\
\midrule

\multicolumn{3}{c}{\cellcolor[gray]{0.95}\textbf{C. Trade \& External Sector}} \\
\midrule
C1 & USTR / Dept. of Commerce & Implementation of new Section 301 tariffs, or export controls (Entity List) affecting key sectors. \\
C2 & Treasury / Fed & Trade-weighted US Dollar Index (DXY) moves $\ge 10\%$ within 3 months. \\
C3 & BEA & Current Account Deficit widens by $>2\%$ of GDP YoY; OR net foreign capital outflows exceed historical 2$\sigma$. \\
\midrule

\multicolumn{3}{c}{\cellcolor[gray]{0.95}\textbf{D. Commodity, Energy \& Supply Chain}} \\
\midrule
D1 & EIA (Energy Info.) & WTI Crude or Henry Hub Natural Gas spot prices change $\ge 30\%$ over 6 months. \\
D2 & USDA / USGS & Key agricultural or metal commodity prices deviate $\ge 25\%$ from 6-month moving average. \\
D3 & Fed NY / Census & Global Supply Chain Pressure Index (GSCPI) exceeds 2 standard deviations. \\
\midrule

\multicolumn{3}{c}{\cellcolor[gray]{0.95}\textbf{E. Real Economy Activity}} \\
\midrule
E1 & Fed Board (G.17) & Industrial Production Index contracts $\ge 3\%$ YoY for 2 consecutive months. \\
E2 & Census Bureau & Retail Sales (ex-auto) contract $\ge 2\%$ YoY; OR Univ. of Michigan Consumer Sentiment drops to bottom 10\%. \\
E3 & FHFA / S\&P CoreLogic & Case-Shiller National Home Price Index turns negative YoY; OR Housing Starts drop $\ge 20\%$ YoY. \\
E4 & BEA / Congress & Tech sector value-add deviates $\ge 2\sigma$ from trend; OR passage of major industrial policy (e.g., CHIPS Act). \\
\midrule

\multicolumn{3}{c}{\cellcolor[gray]{0.95}\textbf{F. Financial Stability \& Credit Cycle}} \\
\midrule
F1 & Fed Board / BIS & Private non-financial sector credit-to-GDP gap exceeds +10\% (Boom) or drops below -5\% (Bust). \\
F2 & FDIC / Fed & NPL ratio for insured institutions rises $\ge 1.0\%$; OR failure/rescue of a SIFI bank. \\
F3 & NYSE / Nasdaq & S\&P 500 or Nasdaq 100 enters Technical Bear Market (drawdown $\ge 20\%$ from peak). \\
\midrule

\multicolumn{3}{c}{\cellcolor[gray]{0.95}\textbf{G. Structural \& Regulatory Policy}} \\
\midrule
G1 & EPA / Congress & Passage of major climate legislation (e.g., IRA subsidies); OR new SEC climate disclosure mandates. \\
G2 & FTC / FCC & Major antitrust lawsuit filed against Big Tech; OR new federal data privacy executive orders. \\
G3 & Congress & Enactment of major reforms to Social Security, Medicare, or Immigration laws. \\
\midrule

\multicolumn{3}{c}{\cellcolor[gray]{0.95}\textbf{H. Labour Market \& Household Sector}} \\
\midrule
H1 & BLS & Unemployment Rate changes $\ge 0.5\%$ (Sahm Rule); OR Non-farm Payrolls deviate $>50$k from consensus. \\
H2 & BEA & Real Disposable Personal Income contracts $\ge 2\%$ YoY. \\
\midrule

\multicolumn{3}{c}{\cellcolor[gray]{0.95}\textbf{I. Geopolitical \& Systemic Shocks}} \\
\midrule
I1 & Dept. of State / OFAC & US becomes party to armed conflict; OR designation of major sanctions on a G20 economy. \\
I2 & FEMA / CDC & Presidential Disaster Declaration for event costing $>\$10$B; OR Nationwide Public Health Emergency declaration. \\
I3 & Treasury / Fed & VIX Index $>35$ combined with net foreign selling of US Treasuries. \\

\end{xltabular}

\begin{xltabular}{\linewidth}{l|p{3.5cm}|X}

\caption{Event grounding standards for the China (CN) market.} \label{tab:grounding_cn} \\

\toprule
\textbf{Code} & \textbf{Authority / Source} & \textbf{Quantifiable Trigger / Definition} \\
\midrule
\endfirsthead

\multicolumn{3}{c}%
{{\bfseries \tablename\ \thetable{} -- continued from previous page}} \\
\toprule
\textbf{Code} & \textbf{Authority / Source} & \textbf{Quantifiable Trigger / Definition} \\
\midrule
\endhead

\midrule
\multicolumn{3}{r}{{Continued on next page...}} \\
\endfoot

\bottomrule
\endlastfoot

\multicolumn{3}{c}{\cellcolor[gray]{0.95}\textbf{A. Monetary \& Financial Conditions}} \\
\midrule
A1 & PBoC (Central Bank) & 7-day Reverse Repo or 1-year MLF rate changes by $\ge 5$ bps; OR Reserve Requirement Ratio (RRR) cut $\ge 25$ bps. \\
A2 & CFETS / NIFC & DR007 (7-day interbank repo rate) deviates $>50$ bps from policy rate for 5+ days; OR PBoC net liquidity injection CNY$500$B/week. \\
A3 & PBoC / NFRA & Adjustment of Macro-Prudential Assessment (MPA) parameters; OR changes to property sector "Three Red Lines" metrics. \\
\midrule

\multicolumn{3}{c}{\cellcolor[gray]{0.95}\textbf{B. Fiscal Policy \& Public Finance}} \\
\midrule
B1 & State Council / MOF & Issuance of Ultra-long Special Sovereign Bonds; OR Local Government Special Bond quota increase $>\text{\yen}1$ Trillion. \\
B2 & MOF / Market Data & 10-year China Government Bond (CGB) yield spikes $\ge 20$ bps in a month; OR major LGFV bond default event. \\
\midrule

\multicolumn{3}{c}{\cellcolor[gray]{0.95}\textbf{C. Trade \& External Sector}} \\
\midrule
C1 & MOFCOM / Customs & Implementation of export controls on strategic materials (e.g., Gallium/Germanium); OR new tariffs on major trading partners. \\
C2 & SAFE / PBoC & USD/CNY Daily Fixing deviates from market close by $>500$ pips (Counter-cyclical factor usage); OR FX Reserves drop $>\$50$B/month. \\
C3 & SAFE & Capital Account net outflows exceed $2\%$ of GDP (annualized); OR major restrictions on cross-border capital flows. \\
\midrule

\multicolumn{3}{c}{\cellcolor[gray]{0.95}\textbf{D. Commodity, Energy \& Supply Chain}} \\
\midrule
D1 & NDRC / NEA & NDRC adjusts guided retail fuel prices; OR thermal coal spot price exceeds regulatory price cap range. \\
D2 & DCE / SHFE & Domestic futures prices for Iron Ore or Rebar deviate $\ge 20\%$ from 6-month MA. \\
D3 & MOT / Caixin & Caixin Manufacturing PMI Suppliers' Delivery Times sub-index drops below 45.0. \\
\midrule

\multicolumn{3}{c}{\cellcolor[gray]{0.95}\textbf{E. Real Economy Activity}} \\
\midrule
E1 & NBS / Caixin & Official Manufacturing PMI or Caixin PMI contracts (<50.0) for 2 consecutive months. \\
E2 & NBS & Retail Sales of Consumer Goods YoY growth turns negative; OR Youth Unemployment Rate (16-24) exceeds 20\%. \\
E3 & NBS / MOHURD & 70-City New Home Price Index declines YoY; OR Top-100 Developer Sales value drops $\ge 20\%$ YoY. \\
E4 & MIIT / NDRC & Launch of major strategic projects (e.g., "East Data West Computing"); OR Strategic Emerging Industries value-add $\ge 2\sigma$ vs trend. \\
\midrule

\multicolumn{3}{c}{\cellcolor[gray]{0.95}\textbf{F. Financial Stability \& Credit Cycle}} \\
\midrule
F1 & PBoC & Total Social Financing (TSF) growth rate gap vs Nominal GDP growth $\ge$ $\pm 5\%$. \\
F2 & NFRA / PBoC & Takeover/Resolution of a medium-sized bank (e.g., Baoshang style event); OR Commercial Bank NPL ratio rises $\ge 0.5\%$. \\
F3 & SSE / SZSE & CSI 300 Index experiences a rapid drawdown $\ge 20\%$ (Technical Bear Market) or triggers trading curbs. \\
\midrule

\multicolumn{3}{c}{\cellcolor[gray]{0.95}\textbf{G. Structural \& Regulatory Policy}} \\
\midrule
G1 & NDRC / MEE & Issuance of "Dual Carbon" (1+N) policy documents; OR launch of new National Carbon Market trading rules. \\
G2 & CAC / SAMR & New anti-monopoly penalties on platform economy firms; OR CAC initiates cybersecurity review on major data handlers. \\
G3 & CPC Central Comm. & "Third Plenum" or "Two Sessions" announces major reforms (e.g., Hukou reform, Common Prosperity initiatives). \\
\midrule

\multicolumn{3}{c}{\cellcolor[gray]{0.95}\textbf{H. Labour Market \& Household Sector}} \\
\midrule
H1 & NBS & Surveyed Urban Unemployment Rate rises $\ge 0.5\%$; OR Migrant Worker population contracts YoY. \\
H2 & NBS / PBoC & Household deposits surge CNY$5$ Trillion YoY (Excess Savings); OR Household Leverage Ratio declines (Deleveraging). \\
\midrule

\multicolumn{3}{c}{\cellcolor[gray]{0.95}\textbf{I. Geopolitical \& Systemic Shocks}} \\
\midrule
I1 & MFA / CMC & Escalation of tensions in Taiwan Strait or South China Sea triggering military exercises; OR foreign sanctions on Chinese entities. \\
I2 & NHC / MEM & Activation of Level-I Public Health Emergency Response; OR natural disaster affecting $>1\%$ of national arable land. \\
I3 & PBoC & Stock Connect / Bond Connect net outflows exceed historical $99$th percentile. \\

\end{xltabular}

\begin{xltabular}{\linewidth}{l|p{3.5cm}|X}
\caption{Event grounding standards for the Japan (JP) market.} \label{tab:grounding_jp} \\

\toprule
\textbf{Code} & \textbf{Authority / Source} & \textbf{Quantifiable Trigger / Definition} \\
\midrule
\endfirsthead

\multicolumn{3}{c}%
{{\bfseries \tablename\ \thetable{} -- continued from previous page}} \\
\toprule
\textbf{Code} & \textbf{Authority / Source} & \textbf{Quantifiable Trigger / Definition} \\
\midrule
\endhead

\midrule
\multicolumn{3}{r}{{Continued on next page...}} \\
\endfoot

\bottomrule
\endlastfoot

\multicolumn{3}{c}{\cellcolor[gray]{0.95}\textbf{A. Monetary \& Financial Conditions}} \\
\midrule
A1 & Bank of Japan (BoJ) & Change in Policy Rate (Uncollateralized Call Rate) $\ge 10$ bps; OR Modification of Yield Curve Control (YCC) band (e.g., widening band). \\
A2 & BoJ / JSDA & 10-year JGB yield breaches the upper limit of the reference range; OR "Rinban" (JGB purchase) operations increase significantly. \\
A3 & BoJ / JFSA & Changes to ETF/J-REIT purchase program guidelines; OR Macro-prudential measures on regional bank real estate lending. \\
\midrule

\multicolumn{3}{c}{\cellcolor[gray]{0.95}\textbf{B. Fiscal Policy \& Public Finance}} \\
\midrule
B1 & Cabinet Office / MoF & Approval of a "Supplementary Budget" (Hosei Yosan) with spending $>\text{\yen}10$ Trillion; OR new economic package announcement. \\
B2 & MoF & JGB Debt Service Cost rises significantly in budget projections; OR Sovereign Rating outlook downgrade due to debt-to-GDP ratio. \\
\midrule

\multicolumn{3}{c}{\cellcolor[gray]{0.95}\textbf{C. Trade \& External Sector}} \\
\midrule
C1 & METI & Imposition of export restrictions on strategic tech materials (e.g., photoresists); OR removal from "White List" of trade partners. \\
C2 & MoF / BoJ & Official FX Intervention confirmed by MoF (buying JPY/selling USD); OR USD/JPY moves $\ge 3\%$ in a single week. \\
C3 & MoF & Current Account Surplus narrows significantly or turns to deficit (due to energy import costs). \\
\midrule

\multicolumn{3}{c}{\cellcolor[gray]{0.95}\textbf{D. Commodity, Energy \& Supply Chain}} \\
\midrule
D1 & METI / TEPCO & Reactivation of Nuclear Power Plants approved; OR Utility companies apply for electricity rate hike $>10\%$. \\
D2 & MAFF & "Food Price Index" within CPI rises $\ge 5\%$ YoY; OR government subsidies for gasoline/wheat prices triggered. \\
D3 & METI / Toyota & Major automaker halts production due to parts shortage; OR disruption in semiconductor supply chain (e.g., Kumamoto fab). \\
\midrule

\multicolumn{3}{c}{\cellcolor[gray]{0.95}\textbf{E. Real Economy Activity}} \\
\midrule
E1 & BoJ (Tankan) & \textbf{Tankan Large Manufacturers DI} drops by $\ge 5$ points; OR Industrial Production contracts $\ge 2\%$ MoM. \\
E2 & Cabinet Office & GDP (Annualized Real Growth) contracts for 2 consecutive quarters (Technical Recession); OR Consumer Confidence Index drops. \\
E3 & MLIT & Land Price Publication (Chika Koji) shows YoY decline in major metropolitan areas; OR Condo prices in Tokyo enter correction. \\
E4 & METI & Announcement of subsidies for strategic sectors (e.g., Rapidus semiconductor project); OR AI strategy guidelines release. \\
\midrule

\multicolumn{3}{c}{\cellcolor[gray]{0.95}\textbf{F. Financial Stability \& Credit Cycle}} \\
\midrule
F1 & BoJ & Bank Lending YoY growth deviates significantly from trend; OR Corporate bankruptcy liabilities surge (Teikoku Databank). \\
F2 & JFSA / BoJ & Regional Bank (Chigin) merger or recapitalization prompted by FSA; OR surfacing of large losses in securities portfolios (e.g., CLOs). \\
F3 & TSE / JPX & Nikkei 225 or TOPIX drops $\ge 20\%$ from peak; OR Volatility Index (JNIV) spikes $>30$. \\
\midrule

\multicolumn{3}{c}{\cellcolor[gray]{0.95}\textbf{G. Structural \& Regulatory Policy}} \\
\midrule
G1 & METI / MoE & GX (Green Transformation) Promotion Act implementation; OR Carbon Pricing (GX League) introduction. \\
G2 & PPC / METI & Enforcement of stricter personal data protection rules; OR new regulations on Generative AI copyright. \\
G3 & Cabinet Office & "New Capitalism" policy initiatives launched; OR major revisions to Labor Standards Act. \\
\midrule

\multicolumn{3}{c}{\cellcolor[gray]{0.95}\textbf{H. Labour Market \& Household Sector}} \\
\midrule
H1 & Rengo / MHLW & \textbf{"Shunto" (Spring Wage Offensive)} agreed wage hike exceeds $3\%$ (or BoJ target level); OR Active Job Openings-to-Applicants Ratio drops. \\
H2 & MHLW / MIC & Real Cash Earnings contract YoY (Wage-Price spiral failure); OR Household Spending (Kakei Chosa) drops YoY. \\
\midrule

\multicolumn{3}{c}{\cellcolor[gray]{0.95}\textbf{I. Geopolitical \& Systemic Shocks}} \\
\midrule
I1 & MoFA / MoD & Major security incidents near Senkaku Islands; OR North Korean missile launch triggering J-Alert system impacting markets. \\
I2 & Cabinet Office & Nankai Trough Earthquake warning issued; OR natural disaster damage estimate $>\text{\yen}1$ Trillion. \\
I3 & BoJ & "Japan Premium" re-emerges in offshore funding markets; OR massive unwinding of Yen Carry Trade. \\

\end{xltabular}

\begin{xltabular}{\linewidth}{l|p{3.5cm}|X}
\caption{Event grounding standards for the United Kingdom (UK) market.} \label{tab:grounding_uk} \\

\toprule
\textbf{Code} & \textbf{Authority / Source} & \textbf{Quantifiable Trigger / Definition} \\
\midrule
\endfirsthead

\multicolumn{3}{c}%
{{\bfseries \tablename\ \thetable{} -- continued from previous page}} \\
\toprule
\textbf{Code} & \textbf{Authority / Source} & \textbf{Quantifiable Trigger / Definition} \\
\midrule
\endhead

\midrule
\multicolumn{3}{r}{{Continued on next page...}} \\
\endfoot

\bottomrule
\endlastfoot

\multicolumn{3}{c}{\cellcolor[gray]{0.95}\textbf{A. Monetary \& Financial Conditions}} \\
\midrule
A1 & BoE (MPC) & Bank Rate change $\ge 25$ bps; OR MPC Vote Split changes significantly (e.g., from 6-3 to 5-4) signaling pivot; OR Active Gilt Sales (QT). \\
A2 & BoE / SONIA & SONIA-Bank Rate spread widens $>20$ bps; OR failure in Gilt Repo market liquidity (Repo rate dislocation). \\
A3 & BoE (FPC) & Adjustment of Countercyclical Capital Buffer (CCyB) rate; OR intervention in LDI (Liability-Driven Investment) fund leverage rules. \\
\midrule

\multicolumn{3}{c}{\cellcolor[gray]{0.95}\textbf{B. Fiscal Policy \& Public Finance}} \\
\midrule
B1 & HM Treasury / OBR & "Autumn Budget" or "Spring Statement" announces discretionary measures $>\text{\pounds}15$B; OR OBR issues warning on fiscal sustainability. \\
B2 & DMO / Markets & 10-year Gilt yield spikes $\ge 30$ bps in a week (Fiscal Tantrum); OR Gilt auction bid-to-cover ratio drops below 1.5. \\
\midrule

\multicolumn{3}{c}{\cellcolor[gray]{0.95}\textbf{C. Trade \& External Sector}} \\
\midrule
C1 & Dept. for Business & Implementation of new post-Brexit border checks (e.g., BTOM) causing delays; OR changes to Windsor Framework rules. \\
C2 & BoE / Markets & GBP/USD (Cable) moves $\ge 3\%$ in a week; OR Sterling Trade-Weighted Index drops significantly (Inflationary devaluation). \\
C3 & ONS & Current Account Deficit exceeds $5\%$ of GDP (Structural vulnerability warning). \\
\midrule

\multicolumn{3}{c}{\cellcolor[gray]{0.95}\textbf{D. Commodity, Energy \& Supply Chain}} \\
\midrule
D1 & OFGEM & OFGEM Energy Price Cap adjustment exceeds $\pm 10\%$ (Direct impact on CPI); OR Govt activates Energy Price Guarantee. \\
D2 & DEFRA / ONS & Food CPI inflation exceeds $10\%$ YoY (Cost of Living Crisis indicator). \\
D3 & CBI / ONS & CBI Industrial Trends Survey "Factors limiting output" (Materials/Labour) spikes above historical average. \\
\midrule

\multicolumn{3}{c}{\cellcolor[gray]{0.95}\textbf{E. Real Economy Activity}} \\
\midrule
E1 & ONS & Monthly GDP (3M/3M) growth turns negative; OR Services PMI drops below 50.0 (Services comprise $\approx 80\%$ of UK economy). \\
E2 & ONS / GfK & GfK Consumer Confidence Index drops below -30; OR Retail Sales volumes contract YoY. \\
E3 & Halifax / Nationwide & Halifax or Nationwide House Price Index falls YoY; OR Mortgage Approvals drop below 50k/month. \\
E4 & DSIT & Announcement of AI Safety Institute initiatives; OR major investments in UK Life Sciences/Tech hubs (e.g., Golden Triangle). \\
\midrule

\multicolumn{3}{c}{\cellcolor[gray]{0.95}\textbf{F. Financial Stability \& Credit Cycle}} \\
\midrule
F1 & BoE & Mortgage lending net flow turns negative; OR Consumer credit growth (credit cards) surges (Distress borrowing). \\
F2 & BoE / PRA & Stress in Challenger Banks; OR rise in corporate insolvencies (Companies House data) exceeding historical averages. \\
F3 & LSE / FTSE & FTSE 250 Index (Domestic proxy) drops $\ge 15\%$; OR widening of Corporate Bond spreads vs Gilts. \\
\midrule

\multicolumn{3}{c}{\cellcolor[gray]{0.95}\textbf{G. Structural \& Regulatory Policy}} \\
\midrule
G1 & DESNZ & Changes to Net Zero 2050 timeline (e.g., delaying ICE car ban); OR changes to Windfall Tax (EGL) on energy firms. \\
G2 & CMA & CMA (Competition and Markets Authority) blocks major tech M\&A; OR new Digital Markets, Competition and Consumers Bill enforcement. \\
G3 & UK Parliament & Passage of major legislation on Renters' Reform or Immigration (Visa salary thresholds). \\
\midrule

\multicolumn{3}{c}{\cellcolor[gray]{0.95}\textbf{H. Labour Market \& Household Sector}} \\
\midrule
H1 & ONS & Average Weekly Earnings (AWE) private sector regular pay growth $>6\%$ (Wage-Price Spiral risk); OR Claimant Count rises. \\
H2 & ONS & Real Household Disposable Income (RHDI) per capita falls for 2 consecutive quarters. \\
\midrule

\multicolumn{3}{c}{\cellcolor[gray]{0.95}\textbf{I. Geopolitical \& Systemic Shocks}} \\
\midrule
I1 & FCDO & UK military involvement in overseas operations; OR major diplomatic rift impacting Trade and Cooperation Agreement (TCA). \\
I2 & Cabinet Office & National Risk Register event activation (e.g., Grid blackout warning); OR Pandemic-level health restrictions. \\
I3 & BoE & "Flash Crash" in Sterling assets; OR systemic margin calls in pension fund LDI strategies. \\

\end{xltabular}

\begin{xltabular}{\linewidth}{l|p{3.5cm}|X}
\caption{Event grounding standards for Germany (DE) market.} \label{tab:grounding_de} \\

\toprule
\textbf{Code} & \textbf{Authority / Source} & \textbf{Quantifiable Trigger / Definition} \\
\midrule
\endfirsthead

\multicolumn{3}{c}%
{{\bfseries \tablename\ \thetable{} -- continued from previous page}} \\
\toprule
\textbf{Code} & \textbf{Authority / Source} & \textbf{Quantifiable Trigger / Definition} \\
\midrule
\endhead

\midrule
\multicolumn{3}{r}{{Continued on next page...}} \\
\endfoot

\bottomrule
\endlastfoot

\multicolumn{3}{c}{\cellcolor[gray]{0.95}\textbf{A. Monetary \& Financial Conditions}} \\
\midrule
A1 & ECB (Governing Council) & ECB Deposit Facility Rate change $\ge 25$ bps; OR ECB announces new asset purchase program (e.g., TPI) to limit spreads. \\
A2 & Bundesbank / ECB & Target2 imbalances for Germany widen significantly; OR Euribor-OIS spread widens $>20$ bps (Interbank stress). \\
A3 & BaFin / Bundesbank & Activation of Countercyclical Capital Buffer (CCyB) for German banks; OR strict LTV caps on residential mortgages. \\
\midrule

\multicolumn{3}{c}{\cellcolor[gray]{0.95}\textbf{B. Fiscal Policy \& Public Finance}} \\
\midrule
B1 & BMF / Bundestag & Suspension of "Schuldenbremse" (Debt Brake) verified by Bundestag; OR Announcement of "Sondervermögen" (Special Fund) $>\text{\euro}50$B. \\
B2 & Finanzagentur & 10-year Bund yield spikes $\ge 30$ bps; OR Bund-BTP (Italy) spread widens $>250$ bps (Eurozone fragmentation risk). \\
\midrule

\multicolumn{3}{c}{\cellcolor[gray]{0.95}\textbf{C. Trade \& External Sector}} \\
\midrule
C1 & BMWK / EU Commission & New EU tariffs on Chinese EVs (affecting German Auto sector); OR Export controls on dual-use goods to major partners. \\
C2 & ECB / Markets & EUR/USD exchange rate moves $\ge 3\%$ in a week; OR Euro Nominal Effective Exchange Rate (NEER) drops significantly. \\
C3 & Destatis / Bundesbank & Current Account Surplus drops below $2\%$ of GDP (Structural loss of competitiveness). \\
\midrule

\multicolumn{3}{c}{\cellcolor[gray]{0.95}\textbf{D. Commodity, Energy \& Supply Chain}} \\
\midrule
D1 & Bundesnetzagentur & TTF Gas Price (Dutch Benchmark) spikes $\ge 30\%$; OR declaration of "Gas Emergency Plan" (Notfallplan Gas) Level 2/3. \\
D2 & Destatis & PPI (Producer Price Index) Energy component rises $\ge 20\%$ YoY. \\
D3 & Ifo Institute & Ifo Survey "Material Shortages" (Materialknappheit) indicator rises above $50\%$ of firms. \\
\midrule

\multicolumn{3}{c}{\cellcolor[gray]{0.95}\textbf{E. Real Economy Activity}} \\
\midrule
E1 & Ifo Institute / Destatis & Ifo Business Climate Index drops for 3 consecutive months; OR Industrial Production (Auto sector) contracts $\ge 5\%$ YoY. \\
E2 & GfK & GfK Consumer Climate index drops below -20 points; OR Retail Sales (Real) contract YoY. \\
E3 & Destatis / Bulwiengesa & Residential Property Price Index contracts YoY; OR Building Permits (Baugenehmigungen) drop $\ge 20\%$ YoY. \\
E4 & BMWK & Announcement of major subsidies for Chip fabs (e.g., Magdeburg Intel plant) or Hydrogen core network ($>\text{\euro}10$B). \\
\midrule

\multicolumn{3}{c}{\cellcolor[gray]{0.95}\textbf{F. Financial Stability \& Credit Cycle}} \\
\midrule
F1 & Bundesbank & Lending to Non-Financial Corporations (NFC) contracts YoY; OR Bank Lending Survey (BLS) shows severe tightening standards. \\
F2 & BaFin & Distress in "Landesbanken" sector; OR Commercial Real Estate (CRE) NPL ratio rises significantly. \\
F3 & Deutsche Börse & DAX 40 index drops $\ge 20\%$ (Bear Market); OR Volatility (VDAX-NEW) spikes $>35$. \\
\midrule

\multicolumn{3}{c}{\cellcolor[gray]{0.95}\textbf{G. Structural \& Regulatory Policy}} \\
\midrule
G1 & BMWK & Implementation of "Heizungsgesetz" (Heating Law/GEG); OR Carbon Price (CO2-Preis) hike $>\text{\euro}10$/ton. \\
G2 & Bundeskartellamt & Federal Cartel Office blocks major merger; OR enforcement of "Digital Services Act" (DSA) penalties on platforms. \\
G3 & Bundestag & Collapse of Coalition Government ("Ampel-Aus"); OR passage of "Wachstumschancengesetz" (Growth Opportunity Act). \\
\midrule

\multicolumn{3}{c}{\cellcolor[gray]{0.95}\textbf{H. Labour Market \& Household Sector}} \\
\midrule
H1 & Bundesagentur für Arbeit & "Kurzarbeit" (Short-time work) notifications exceed 100k/month; OR Unemployment Rate rises $\ge 0.5\%$. \\
H2 & Destatis & Real Wages (Reallöhne) contract YoY for 2 consecutive quarters. \\
\midrule

\multicolumn{3}{c}{\cellcolor[gray]{0.95}\textbf{I. Geopolitical \& Systemic Shocks}} \\
\midrule
I1 & Auswärtiges Amt & Major disruption to Nord Stream or critical energy infrastructure; OR Germany increases Defense Fund ($>\text{\euro}100$B). \\
I2 & BBK & National warning day activation for critical infrastructure failure; OR Rhine water levels drop below "Kaub" critical mark (halting shipping). \\
I3 & ECB / Bundesbank & Spreads between Core (Bund) and Periphery (BTP) widen $>250$ bps (Fragmentation risk) triggering ECB intervention. \\

\end{xltabular}

\begin{xltabular}{\linewidth}{l|p{3.5cm}|X}
\caption{Event grounding standards for the France (FR) market.} \label{tab:grounding_fr} \\

\toprule
\textbf{Code} & \textbf{Authority / Source} & \textbf{Quantifiable Trigger / Definition} \\
\midrule
\endfirsthead

\multicolumn{3}{c}%
{{\bfseries \tablename\ \thetable{} -- continued from previous page}} \\
\toprule
\textbf{Code} & \textbf{Authority / Source} & \textbf{Quantifiable Trigger / Definition} \\
\midrule
\endhead

\midrule
\multicolumn{3}{r}{{Continued on next page...}} \\
\endfoot

\bottomrule
\endlastfoot

\multicolumn{3}{c}{\cellcolor[gray]{0.95}\textbf{A. Monetary \& Financial Conditions}} \\
\midrule
A1 & ECB / BdF & ECB Deposit Facility Rate change $\ge 25$ bps; OR Banque de France Governor speech signaling deviation from consensus. \\
A2 & BdF / Euronext & 3-month Euribor spread vs OIS widens $>20$ bps; OR Repo market fragmentation for French collateral. \\
A3 & HCSF / BdF & HCSF (High Council for Financial Stability) adjusts countercyclical buffer; OR enforcement of strict 35\% Debt-Service-to-Income (DSTI) cap on mortgages. \\
\midrule

\multicolumn{3}{c}{\cellcolor[gray]{0.95}\textbf{B. Fiscal Policy \& Public Finance}} \\
\midrule
B1 & Ministry of Economy / Parliament & Passage of "Projet de loi de finances" (PLF) via Article 49.3 (forcing adoption without vote); OR Deficit exceeds 3\% Maastricht limit triggering EU Excessive Deficit Procedure. \\
B2 & AFT / Markets & 10-year OAT yield spikes $\ge 30$ bps; OR OAT-Bund spread widens $>50$ bps (signaling sovereign risk premium). \\
\midrule

\multicolumn{3}{c}{\cellcolor[gray]{0.95}\textbf{C. Trade \& External Sector}} \\
\midrule
C1 & Customs / EU & New EU Carbon Border Adjustment Mechanism (CBAM) implementation affecting French industry; OR trade disputes on Luxury Goods sector. \\
C2 & BdF / Markets & EUR/USD volatility $>15\%$ annualized; OR Real Effective Exchange Rate (REER) appreciation hurting export competitiveness. \\
C3 & BdF & Current Account Deficit widens $>\text{\euro}10$B in a quarter; OR deterioration in Trade Balance due to energy imports. \\
\midrule

\multicolumn{3}{c}{\cellcolor[gray]{0.95}\textbf{D. Commodity, Energy \& Supply Chain}} \\
\midrule
D1 & CRE / EDF & EDF Nuclear Output drops below 280 TWh/year (historical low); OR Government adjusts "Bouclier tarifaire" (Tariff Shield) cap on electricity prices. \\
D2 & INSEE & Food CPI inflation exceeds $10\%$ YoY (Panier anti-inflation monitoring). \\
D3 & BdF / INSEE & Business Sentiment (Climat des affaires) "Supply Difficulties" sub-index rises significantly. \\
\midrule

\multicolumn{3}{c}{\cellcolor[gray]{0.95}\textbf{E. Real Economy Activity}} \\
\midrule
E1 & INSEE & Manufacturing Output contracts $\ge 1\%$ MoM; OR Business Climate Index (Climat des affaires) drops below 100 long-term average. \\
E2 & INSEE & Consumer Confidence (Confiance des ménages) drops below 85; OR Household Consumption of goods contracts YoY. \\
E3 & Notaires de France / INSEE & Index of Existing Home Prices falls YoY; OR Housing Starts (Mises en chantier) drop $\ge 15\%$ YoY. \\
E4 & Ministry of Economy & "France 2030" investment plan disbursements acceleration; OR major subsidies for "Gigafactories" (Batteries) in Northern France. \\
\midrule

\multicolumn{3}{c}{\cellcolor[gray]{0.95}\textbf{F. Financial Stability \& Credit Cycle}} \\
\midrule
F1 & BdF & Credit to Non-Financial Corporations growth slows to $<2\%$ YoY; OR rise in "Prêts Garantis par l’État" (PGE) defaults. \\
F2 & ACPR / BdF & Solvency ratio of major Bancassurance groups drops; OR rise in Life Insurance (Assurance Vie) withdrawals. \\
F3 & Euronext Paris & CAC 40 Index drops $\ge 20\%$ (Bear Market); OR Luxury Sector sub-index (LVMH, Hermes, Kering) corrects $\ge 15\%$. \\
\midrule

\multicolumn{3}{c}{\cellcolor[gray]{0.95}\textbf{G. Structural \& Regulatory Policy}} \\
\midrule
G1 & Ministry of Ecology & New "DPE" (Energy Performance Diagnosis) bans on renting G-rated housing; OR Carbon Tax increase. \\
G2 & CNIL & CNIL fines major tech firm for GDPR violation; OR new "Influencer Law" regulation enforcement. \\
G3 & Parliament / President & Passage of Pension Reform (Réforme des retraites) raising retirement age; OR Unemployment Insurance reform decrees. \\
\midrule

\multicolumn{3}{c}{\cellcolor[gray]{0.95}\textbf{H. Labour Market \& Household Sector}} \\
\midrule
H1 & DARES / Unions & General Strike (Grève générale) disrupting Transport/Refineries for $>3$ days; OR Private Sector Payrolls (Emploi salarié) contract. \\
H2 & INSEE & Purchasing Power (Pouvoir d'achat) per unit contracts YoY; OR SMIC (Minimum Wage) automatic inflation adjustment $>2\%$. \\
\midrule

\multicolumn{3}{c}{\cellcolor[gray]{0.95}\textbf{I. Geopolitical \& Systemic Shocks}} \\
\midrule
I1 & Ministry of Foreign Affairs & Direct French military intervention (e.g., Sahel, Eastern Europe); OR Terror Alert Level raised to "Urgence Attentat". \\
I2 & Ministry of Interior & Civil Unrest (e.g., "Gilets Jaunes" or 2023 Riots) causing nationwide damage $>\text{\euro}1$B; OR Drought restrictions impacting agriculture. \\
I3 & BdF / ECB & OAT-Bund Spread widening $>80$ bps triggering ECB TPI activation. \\

\end{xltabular}

\begin{xltabular}{\linewidth}{l|p{3.5cm}|X}
\caption{Event grounding standards for Singapore (SG) market.} \label{tab:grounding_sg} \\

\toprule
\textbf{Code} & \textbf{Authority / Source} & \textbf{Quantifiable Trigger / Definition} \\
\midrule
\endfirsthead

\multicolumn{3}{c}%
{{\bfseries \tablename\ \thetable{} -- continued from previous page}} \\
\toprule
\textbf{Code} & \textbf{Authority / Source} & \textbf{Quantifiable Trigger / Definition} \\
\midrule
\endhead

\midrule
\multicolumn{3}{r}{{Continued on next page...}} \\
\endfoot

\bottomrule
\endlastfoot

\multicolumn{3}{c}{\cellcolor[gray]{0.95}\textbf{A. Monetary \& Financial Conditions}} \\
\midrule
A1 & MAS & MAS tightens S\$NEER Policy Band (e.g., steepening slope, widening width, or re-centering upwards); OR Off-cycle MPS announcement. \\
A2 & MAS / ABS & SORA (Singapore Overnight Rate Average) spikes $>50$ bps in a week; OR SGD liquidity facility usage surges. \\
A3 & MAS & Implementation of tighter Total Debt Servicing Ratio (TDSR) threshold (e.g., from 60\% to 55\%); OR reduction in LTV limits. \\
\midrule

\multicolumn{3}{c}{\cellcolor[gray]{0.95}\textbf{B. Fiscal Policy \& Public Finance}} \\
\midrule
B1 & MOF & "Budget Statement" announces major "Cost of Living" Support Package ($>\text{S\$}5$B); OR Distribution of CDC/GST Vouchers. \\
B2 & MOF / Yield Curve & \textbf{SGS} (Singapore Govt Securities) yield curve inverts significantly; OR Net Investment Returns Contribution (NIRC) projected to drop. \\
\midrule

\multicolumn{3}{c}{\cellcolor[gray]{0.95}\textbf{C. Trade \& External Sector}} \\
\midrule
C1 & Enterprise SG & NODX (Non-Oil Domestic Exports) contracts $\ge 10\%$ YoY (Tech cycle downturn); OR major supply chain disruption affecting Port of Singapore. \\
C2 & MAS & S\$NEER hits the top or bottom of the policy band (Implicit intervention zone); OR SIBOR/SORA spread vs USD rates widens sharply. \\
C3 & MTI & Balance of Payments deficit driven by capital outflows; OR significant drop in Official Foreign Reserves. \\
\midrule

\multicolumn{3}{c}{\cellcolor[gray]{0.95}\textbf{D. Commodity, Energy \& Supply Chain}} \\
\midrule
D1 & SP Group / EMA & Electricity Tariff adjustment $\ge 10\%$ QoQ; OR Energy Market Authority (EMA) activates fuel stockpile. \\
D2 & MTI / DOS & MAS Core Inflation exceeds $5\%$ YoY (Triggering cost-of-living policy response). \\
D3 & PSA / MPA & Singapore Port Container Throughput contracts YoY; OR Bunker fuel prices spike disrupting shipping hub status. \\
\midrule

\multicolumn{3}{c}{\cellcolor[gray]{0.95}\textbf{E. Real Economy Activity}} \\
\midrule
E1 & EDB & Industrial Production (Electronics Cluster) contracts $\ge 10\%$ YoY; OR PMI (SIPMM) drops below 50.0. \\
E2 & DOS & Retail Sales (excluding Motor Vehicles) contracts YoY; OR Food \& Beverage Services Index drops significanty. \\
E3 & URA / HDB & URA Private Property Price Index rises $>2\%$ QoQ (Triggering Cooling Measures); OR HDB Resale Price Index surges. \\
E4 & EDB / MAS & Announcement of major FDI in Semiconductor/Biomed (e.g., GlobalFoundries expansion); OR new FinTech regulatory sandbox grants. \\
\midrule

\multicolumn{3}{c}{\cellcolor[gray]{0.95}\textbf{F. Financial Stability \& Credit Cycle}} \\
\midrule
F1 & MAS & Credit-to-GDP gap exceeds 10\%; OR DBU (Domestic Banking Unit) non-bank loans growth turns negative. \\
F2 & MAS & NPL Ratio of Local Banks (DBS/OCBC/UOB) rises $\ge 0.5\%$; OR significant exposure to regional defaults (e.g., China property). \\
F3 & SGX & Straits Times Index (STI) drops $\ge 15\%$; OR daily securities market turnover drops below 10-year average. \\
\midrule

\multicolumn{3}{c}{\cellcolor[gray]{0.95}\textbf{G. Structural \& Regulatory Policy}} \\
\midrule
G1 & MSE / NCCS & Carbon Tax hike (e.g., reaching S\$25/tonne); OR implementation of mandatory climate reporting for ListCos. \\
G2 & MAS & Issuance of updated Technology Risk Management (TRM) Guidelines; OR major enforcement on digital payment token service providers. \\
G3 & Parliament & Changes to CPF (Central Provident Fund) contribution rates or retirement age; OR tightening of Employment Pass (EP) criteria. \\
\midrule

\multicolumn{3}{c}{\cellcolor[gray]{0.95}\textbf{H. Labour Market \& Household Sector}} \\
\midrule
H1 & MOM & Retrenchments exceed 5,000/quarter; OR Unemployment Rate rises $\ge 0.5\%$. \\
H2 & MOM / NWC & Real Median Income Growth turns negative; OR National Wages Council (NWC) recommends wage freeze. \\
\midrule

\multicolumn{3}{c}{\cellcolor[gray]{0.95}\textbf{I. Geopolitical \& Systemic Shocks}} \\
\midrule
I1 & MFA & Major geopolitical tension in Malacca Strait affecting trade flows; OR sanctions impacting Singapore's role as a transshipment hub. \\
I2 & MOH / NEA & DORSCON Level raised to Orange/Red (Pandemic); OR "Haze" PSI index exceeds 300 (Hazardous) disrupting economy. \\
I3 & MAS & S\$NEER falls below policy band midpoint despite tightening (Loss of confidence); OR spike in SGD funding costs. \\

\end{xltabular}

\subsection{Non-recurrent Corporate Event Types}

To ensure our taxonomy is grounded in established financial industry standards rather than subjective definitions, we adopted the \textit{Key Developments} framework from \citet{sp_key_dev}. This database is widely recognized by professional analysts for tracking material corporate episodes and serves as the "gold standard" for event categorization. 

\noindent While the original S\&P schema comprises over 100 granular event types, we curated a streamlined subset of 70 event types (Table~\ref{tab:corporate_events}) specifically optimized for the LLM forecasting task. Our selection process was guided by two principles: (1) \textbf{Materiality and Signal Value:} We filtered out low-impact administrative events (e.g., routine conference presentations, minor website updates) that typically do not trigger significant market reactions or require complex reasoning. (2) \textbf{Semantic Distinctiveness:} We consolidated highly overlapping categories to create a MECE framework suitable for classification tasks.

This curation ensures that the benchmark evaluates the model's ability to anticipate "market-moving" events rather than getting distracted by administrative noise.

\small
\setlength{\tabcolsep}{1.5mm} 
\renewcommand{\arraystretch}{1.1} 

\begin{xltabular}{\linewidth}{c|p{3.4cm}|X||c|p{3.4cm}|X}

\caption{Corporate event types for non-recurrent tasks (70 types across 8 categories).} \label{tab:corporate_events} \\

\toprule
\textbf{No.} & \textbf{Event Type} & \textbf{Description} & \textbf{No.} & \textbf{Event Type} & \textbf{Description} \\
\midrule
\endfirsthead

\multicolumn{6}{c}{{\bfseries \tablename\ \thetable{} -- continued from previous page}} \\
\toprule
\textbf{No.} & \textbf{Event Type} & \textbf{Description} & \textbf{No.} & \textbf{Event Type} & \textbf{Description} \\
\midrule
\endhead

\midrule
\multicolumn{6}{r}{{Continued on next page...}} \\
\endfoot

\bottomrule
\endlastfoot

\multicolumn{3}{c||}{\cellcolor[gray]{0.9}\textbf{Corporate Structure (8)}} & \multicolumn{3}{c}{\cellcolor[gray]{0.9}\textbf{Financial - Results \& Guidance (9)}} \\
\midrule
1 & Executive Changes - CEO & CEO position change & 9 & Announcements of Earnings & Periodic earnings release \\
2 & Executive Changes - CFO & CFO position change & 10 & Announcement of Op. Results & Periodic operating results \\
3 & Executive/Board - Other & Other leadership changes & 11 & Announcements of Sales & Non-earnings results \\
4 & Name Changes & Company name change & 12 & Guidance - Lowered & Lower expected results \\
5 & Address Changes & HQ relocation & 13 & Guidance - Raised & Higher expected results \\
6 & Legal Structure Changes & Legal structure change & 14 & Guidance - New/Confirmed & New or confirmed guidance \\
7 & Business Reorganizations & Unit combining/separating & 15 & Guidance - Unusual Events & Unusual event impact \\
8 & Fiscal Year End Changes & Reporting period change & 16 & Restatements of Results & Correction of results \\
 & & & 17 & Impairments/Write Offs & Asset value reduction \\

\midrule
\multicolumn{3}{c||}{\cellcolor[gray]{0.9}\textbf{Financial - Dividends (6)}} & \multicolumn{3}{c}{\cellcolor[gray]{0.9}\textbf{Financial - Capital \& Financing (10)}} \\
\midrule
18 & Dividend Affirmations & Dividend reiteration & 24 & Buyback Announcements & Repurchase authorized \\
19 & Dividend Increases & Higher dividend & 25 & Buyback Closings & Repurchase concluded \\
20 & Dividend Decreases & Lower dividend & 26 & Buyback Cancellations & Buyback cancelled \\
21 & Dividend Cancellation & Dividend suspended & 27 & Buyback - Plan Changes & Buyback terms modified \\
22 & Dividend Initiation & Maiden dividend & 28 & Stock Splits \& Dividends & Shares increased \\
23 & Special Dividend & Non-recurring distribution & 29 & Debt Financing Related & Debt instruments raised \\
 & & & 30 & Private Placements & Institutional sales \\
 & & & 31 & IPOs & Initial Public Offering \\
 & & & 32 & Follow-on Equity Offerings & Post-IPO equity \\
 & & & 33 & Fixed Income Offerings & Bond/note offering \\

\midrule
\multicolumn{3}{c||}{\cellcolor[gray]{0.9}\textbf{Strategic Actions (10)}} & \multicolumn{3}{c}{\cellcolor[gray]{0.9}\textbf{M\&A Transactions (5)}} \\
\midrule
34 & Business Expansions & Internal growth & 43 & M\&A Rumors \& Discussions & Pre-agreement talks \\
35 & Discontinued Operations & Line/facility closure & 44 & M\&A Announcements & Deal announced \\
36 & Strategic Alliances & Collaboration agreement & 45 & M\&A Closings & Deal finalized \\
37 & Client Announcements & Client relationship change & 46 & M\&A Cancellations & Deal cancelled \\
38 & Product Announcements & Product intro/change & 47 & Spin-Off/Split-Off & Spin-off transaction \\
\cmidrule{4-6}
39 & Strategic Alternatives & Evaluating options & \multicolumn{3}{c}{\cellcolor[gray]{0.9}\textbf{Market Events (5)}} \\
\cmidrule{4-6}
40 & Seeking to Sell/Divest & Sell-off opportunities & 48 & Delistings & Exchange removal \\
41 & Seeking Acquisitions & Acquire interests & 49 & Exchange Changes & Exchange transfer \\
42 & Seeking Financing & Equity/venture funding & 50 & Ticker Changes & Ticker symbol change \\
 & & & 51 & Index Constituent Drops & Index removal \\
 & & & 52 & Index Constituent Adds & Index addition \\

\midrule
\multicolumn{3}{c||}{\cellcolor[gray]{0.9}\textbf{Regulatory Compliance (9)}} & \multicolumn{3}{c}{\cellcolor[gray]{0.9}\textbf{Distress Indicators (7)}} \\
\midrule
53 & Regulatory Inquiries & SEC/regulatory inquiry & 62 & Debt Defaults & Obligation failure \\
54 & Regulatory - Regulations & New rules issued & 63 & Bankruptcy - Filing & Reorganize/liquidate filing \\
55 & Regulatory - Compliance & Approvals/licenses & 64 & Bankruptcy - Conclusion & Petition dismissed \\
56 & Regulatory - Enforcement & Fines/penalties & 65 & Bankruptcy - Emergence & Exit from bankruptcy \\
57 & Lawsuits \& Legal Issues & Legal actions & 66 & Bankruptcy - Other & Other bankruptcy motions \\
58 & Delayed SEC Filings & Late document filing & 67 & Bankruptcy - Asset Sale & Liquidation/asset sale \\
59 & Delayed Earnings & Postponed release & 68 & Halt/Resume Operations & Unusual event halt \\
\cmidrule{4-6}
60 & Auditor Changes & Auditor change & \multicolumn{3}{c}{\cellcolor[gray]{0.9}\textbf{Operational Changes (2)}} \\
\cmidrule{4-6}
61 & Auditor Going Concern & Viability concerns & 69 & Labor Announcements & Labor force news \\
 & & & 70 & Bylaws/Rules Changes & Governance rule changes \\

\end{xltabular}

\section{Non-Recurrent Task Generation Pipeline}
\label{app:nonrecurrent_pipeline}

The Non-Recurrent Task Generation Agent employs a three-stage LLM-based pipeline. 
In \textit{Signal Detection}, search-augmented LLMs scan real-time news streams to identify candidate events with binary outcomes and genuine uncertainty, prioritizing policy announcements, regulatory rulings, and personnel changes while excluding routine scheduled releases.
In \textit{Relevance Assessment}, sub-agents investigate each candidate by analyzing specific unknowns (e.g., debated thresholds, contested wording), conflicting signals across sources, and risk factors such as potential delays, assigning uncertainty scores based on source disagreement.
In \textit{Question Generation}, the agent synthesizes research reports to formulate questions about \textit{marginal content} rather than mere occurrence with automated verification ensuring binary answerability and genuine uncertainty.

\section{Ground Truth Acquisition for Non-Recurrent Tasks}
\label{app:groundtruth_nonrecurrent}

Ground truth for non-recurrent tasks combines automated evidence aggregation with expert validation.
A search-augmented LLM agent executes a three-phase protocol: (1) searching official sources and major financial news, (2) verifying numeric thresholds and event definitions, and (3) checking for denials or contradictory evidence.
An LLM adjudicator synthesizes findings into a binary determination, classifying events as \texttt{YES} only when confirmed by authoritative sources; ambiguous signals or absence of evidence result in \texttt{NO}.
To ensure gold-standard quality, \textbf{all (100\%)} automated proposals undergo mandatory expert review. Specialists verify evidence against official sources for final adjudication, achieving 95\% inter-annotator agreement.

\clearpage
\section{Statistics and Analysis}\label{sec:statistics}

Beyond the construction pipeline and quality control mechanisms, we provide detailed statistics to characterize the scale and distribution of the benchmark. Table~\ref{tab:statistics} presents the basic statistics across the dual-track taxonomy.

\begin{table}[t]
\centering
\caption{Basic statistics of \bench{}. W = Weekly, M = Monthly, Q = Quarterly.}
\label{tab:statistics}
\small 
\renewcommand{\arraystretch}{1.2} 

\begin{tabular*}{\textwidth}{@{\extracolsep{\fill}} l cccc @{}}
\toprule
& \multicolumn{2}{c}{\textbf{Recurrent}} & \multicolumn{2}{c}{\textbf{Non-Recurrent}} \\
\cmidrule(lr){2-3} \cmidrule(lr){4-5}
\textbf{Specification} & \textbf{Macro} & \textbf{Corporate} & \textbf{Macro} & \textbf{Corporate} \\
\midrule

\# Corporates / Economies & 8 & 1,314 & 8 & 1,314 \\
\# Metrics / Event Types & 96 & 121 & 208 & 70 \\
Forecast Horizon & W / M / Q & Q & W / M / Q & W / M / Q \\

\midrule

Avg. Weekly Tasks & 30 & 72 & 16 & 82 \\
Total Tasks       & 296 & 723 & 128 & 247 \\

\bottomrule
\end{tabular*}

\end{table}

\noindent \textbf{Comprehensive Task Coverage.}
The benchmark covers 1,314 companies spanning major financial centers across North America, Europe, and Asia-Pacific, selected from 8 economies and 9 indices. Recurrent tasks target 96 macro indicators and 121 corporate metrics indicators through numerical forecasting; non-recurrent tasks address 208 economy-specific macro event specifications and 70 corporate event types through binary prediction.

\noindent \textbf{Generation Status.}
Since 27 October 2025, the benchmark has generated 1,394 total tasks across 10 weekly batches, averaging 139 new tasks per week. As shown in Table~\ref{tab:statistics}, recurrent tasks total 1,019 with 30 macro tasks and 72 corporate generated weekly on average. Non-recurrent tasks total 375, averaging 16 macro tasks and 82 corporate per week. Ground truth has been obtained for all 1,394 tasks, ensuring complete data availability for model evaluation.

\noindent \textbf{Multi-Frequency Forecast Horizons.}
Three forecast horizons enable differentiated temporal assessment: weekly horizons evaluate short-term responsiveness, monthly horizons assess medium-term trend integration, and quarterly horizons test long-horizon forecasting. Recurrent corporate tasks focus exclusively on quarterly forecasting aligned with earnings cycles, while recurrent macro tasks and all non-recurrent tasks support weekly, monthly, and quarterly horizons.

\noindent \textbf{Cross-Market Task Distribution}
\label{app:market_distribution}
Table~\ref{tab:market_distribution} presents the detailed task distribution across 8 economies and four task categories.

\begin{table}[ht]
\centering
\caption{Task distribution across 8 economies.}
\label{tab:market_distribution}
\small 
\renewcommand{\arraystretch}{1.2} 

\begin{tabular*}{\textwidth}{@{\extracolsep{\fill}} l cccccccc c @{}}
\toprule
\multirow{2}{*}{\textbf{Task Category}} 
 & \includegraphics[width=0.35cm]{arxiv/icons/US.png} 
 & \includegraphics[width=0.35cm]{arxiv/icons/CN.png} 
 & \includegraphics[width=0.35cm]{arxiv/icons/HK.png} 
 & \includegraphics[width=0.35cm]{arxiv/icons/JP.png} 
 & \includegraphics[width=0.35cm]{arxiv/icons/UK.png} 
 & \includegraphics[width=0.35cm]{arxiv/icons/DE.png} 
 & \includegraphics[width=0.35cm]{arxiv/icons/FR.png} 
 & \includegraphics[width=0.35cm]{arxiv/icons/SG.png} 
 & \multirow{2}{*}{\textbf{Total}} \\
 & \textbf{US} & \textbf{CN} & \textbf{HK} & \textbf{JP} & \textbf{UK} & \textbf{DE} & \textbf{FR} & \textbf{SG} & \\
\midrule

Recurrent Macro         & 30  & 40  & 39  & 39  & 40  & 38  & 40  & 30  & 296 \\
Recurrent Corporate     & 413 & 68  & 37  & 141 & 25  & 1   & 24  & 14  & 723 \\
Non-Recurrent Macro     & 16  & 16  & 16  & 16  & 16  & 16  & 16  & 16  & 128 \\
Non-Recurrent Corporate & 57  & 25  & 26  & 30  & 30  & 29  & 30  & 20  & 247 \\

\midrule
\textbf{Total}          & \textbf{516} & \textbf{149} & \textbf{118} & \textbf{226} & \textbf{111} & \textbf{84} & \textbf{110} & \textbf{80} & \textbf{1,394} \\
\bottomrule
\end{tabular*}
\end{table}

The distribution reveals distinct patterns across task categories. Recurrent macro tasks maintain relatively uniform distribution across economies through systematic coverage of scheduled indicator releases. In contrast, recurrent corporate tasks exhibit the largest cross-market variation, reflecting differences in index constituent counts and earnings release timing during our evaluation period. Non-recurrent macro tasks achieve uniformity with exactly 16 tasks per economy by design, while corporate tasks show moderate variation driven by news coverage density. By geographic region, Asia-Pacific accounts for 41\% of total tasks, Americas for 37\%, and Europe for 22\%. This balanced geographic composition is critical for ensuring that the model's performance is evaluated against diverse market microstructures, regulatory environments, and cultural contexts, rather than being overfitted to a single dominant economy like the US.

\section{Model Settings}
\label{sec:model_settings}
Table~\ref{tab:model_ids_settings} presents the method configurations evaluated in our benchmark. All unspecified settings remained at their default values.

\section{Error Case Study}
\label{sec:error_case_study}

Table~\ref{tab:error-analysis} breaks down the most common error modes of
\textbf{Perplexity Sonar Deep Research} across scenarios (Non-Recurrent vs.\ Recurrent)
and levels (Corporate vs.\ Macro). For each scenario--level group, we report the total
number of tasks, the number of incorrect predictions, and the resulting error rate.
To illustrate typical failures, we also provide a representative example from the most
frequent error subtype within each group.

\section{Samples of Model Forecasting}
\label{sec:sample_forecasting}

\begin{figure*}
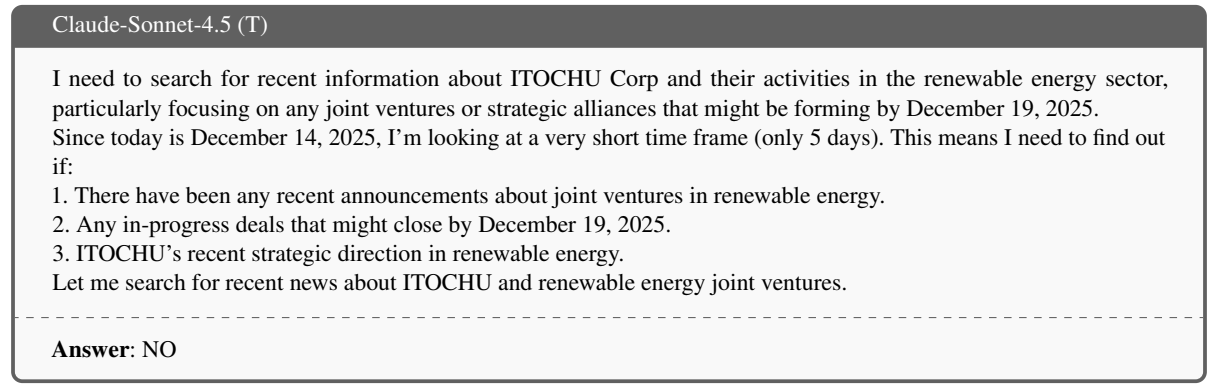


\begin{tcolorbox}[colback=gray!5!white, colframe=gray!75!black, title=Claude-Sonnet-4.5 (T)]

I need to search for recent information about ITOCHU Corp and their activities in the renewable energy sector, particularly focusing on any joint ventures or strategic alliances that might be forming by December 19, 2025.

Since today is December 14, 2025, I'm looking at a very short time frame (only 5 days). This means I need to find out if:

1. There have been any recent announcements about joint ventures in renewable energy.

2. Any in-progress deals that might close by December 19, 2025.

3. ITOCHU's recent strategic direction in renewable energy.

Let me search for recent news about ITOCHU and renewable energy joint ventures.

\tcblower
\textbf{Answer}: NO
\end{tcolorbox}
\caption{Model Reasoning by Claude-Sonnet-4.5 (T) for the Corporate Non-Recurrent Task. The model's answer is NO.}
\label{fig:claude-t-corp}
\end{figure*}

The prompt configuration for a sample corporate non-recurrent task is illustrated in Figure~\ref{fig:prompt_corp}. Subsequent figures delineate the reasoning and answers generated by the evaluated models for the task: OpenAI o3-deep-research (Figure~\ref{fig:gpt-dr-corp}), Perplexity Sonar Deep Research (Figure~\ref{fig:perplexity-dr-corp}), Tongyi Deep Research (Figure~\ref{fig:tongyi-dr-corp}), OpenAI GPT-5 (T+S) (Figure~\ref{fig:gpt-t+s-corp}), and Claude-Sonnete-4.5 (T) (Figure~\ref{fig:claude-t-corp}).

\clearpage

\small
\renewcommand{\arraystretch}{1.1} 
\setlength{\tabcolsep}{3mm}

\begin{xltabular}{\linewidth}{p{3.2cm}|p{5cm}|>{\raggedright\arraybackslash}X}

\caption{Correspondence between benchmark aliases, API identifiers, and API settings} \label{tab:model_ids_settings} \\
\toprule
\textbf{Benchmark Alias} & \textbf{API Identifier} & \textbf{API Setting} \\
\midrule
\endfirsthead

\multicolumn{3}{c}{{\bfseries \tablename\ \thetable{} -- continued from previous page}} \\
\toprule
\textbf{Benchmark Alias} & \textbf{API Identifier} & \textbf{API Setting} \\
\midrule
\endhead

\midrule
\multicolumn{3}{r}{{Continued on next page...}} \\
\endfoot

\bottomrule
\endlastfoot

\multicolumn{3}{c}{\cellcolor[gray]{0.95}\textbf{LLM (Thinking)}} \\
\midrule
Gemini 2.5 Pro (T) & gemini-2.5-pro-preview-05-06 & \seqsplit{thinking\_budget=-1} \\
Deepseek-v3.2 (T) & deepseek-v3.2-exp & \seqsplit{reasoning.enabled=True} \\
Claude-Sonnet-4.5 (T) & claude-sonnet-4-5-20250929 & \seqsplit{thinking.type=enabled, thinking.budget\_tokens=10000} \\
Grok 4 (T) & grok-4-0709 & all defaults \\
OpenAI GPT-5 (T) & gpt-5-2025-08-07 & \seqsplit{reasoning.effort=high} \\
\midrule

\multicolumn{3}{c}{\cellcolor[gray]{0.95}\textbf{LLM (Thinking + Search)}} \\
\midrule
Gemini 2.5 Pro (T+S) & gemini-2.5-pro-preview-05-06 & \seqsplit{thinking\_budget=-1, tools=[google\_search]} \\
Deepseek-v3.2 (T+S) & deepseek-v3.2-exp & \seqsplit{reasoning.enabled=True, plugins=[exa(max\_results=8)]} \\
Claude-Sonnet-4.5 (T+S) & claude-sonnet-4-5-20250929 & \seqsplit{thinking.type=enabled, thinking.budget\_tokens=10000, tools=[web\_search\_20250305]} \\
Grok 4 (T+S) & grok-4-0709 & \seqsplit{search\_parameters.mode=on} \\
OpenAI GPT-5 (T+S) & gpt-5-2025-08-07 & \seqsplit{reasoning.effort=medium, tools=[web\_search]} \\
\midrule

\multicolumn{3}{c}{\cellcolor[gray]{0.95}\textbf{Deep Research}} \\
\midrule
Perplexity Sonar Deep Research & sonar-deep-research & \seqsplit{reasoning.effort=high} \\
Tongyi Deep Research & tongyi-deepresearch-30b-a3b & \seqsplit{temperature=0.6, top\_p=0.95, presence\_penalty=1.1} \\
OpenAI o3-deep-research & o3-deep-research-2025-06-26 & \seqsplit{tools=[web\_search\_preview, code\_interpreter]} \\

\end{xltabular}

\begin{table*}[!t]
\centering
\small
\caption{Error analysis for \textbf{Perplexity Sonar Deep Research}. In Representative Error Case, we show the most common error subtype (Type/Metric/Indicator) and its share of that group’s errors (parenthesized \%), followed by an example question (Q), gold answer, and model prediction.}
\renewcommand{\arraystretch}{1.25} 
\setlength{\tabcolsep}{5pt} 

\begin{tabularx}{\textwidth}{@{} ll r r r >{\raggedright\arraybackslash}X @{}}
\toprule
\textbf{Scenario} & \textbf{Level} & \textbf{\# Tasks} & \textbf{\# Errors} & \textbf{\% Error} & \textbf{Representative Error Case} \\
\midrule

\multirow{3}{*}[-2pt]{Non-Recurrent} & \multirow{3}{*}[-2pt]{Corporate} & \multirow{3}{*}[-2pt]{247} & \multirow{3}{*}[-2pt]{50} & \multirow{3}{*}[-2pt]{20.24} & 
    {\footnotesize \textbf{Type:} Corporate Event (32.0\%)} \newline
    \textbf{Q:} Is it probable that Air Liquide SA will announce a new production facility in North America by 2025-12-19? \newline
    \textbf{Gold:} \texttt{YES} \quad \textbf{Pred:} \textcolor{red}{\texttt{NO}} \\
\midrule

\multirow{7}{*}[3pt]{Non-Recurrent} & \multirow{7}{*}[3pt]{Macro} & \multirow{7}{*}[3pt]{128} & \multirow{7}{*}[3pt]{34} & \multirow{7}{*}[3pt]{26.56} & 
    {\footnotesize \textbf{Type:} --} \newline
    \textbf{Q:} Will UK Chancellor Rachel Reeves, in the Autumn Budget statement on November 26, 2025, announce at least one specific new tax increase measure (beyond what was already announced or leaked before November 17) that was not explicitly mentioned in official government communications prior to November 17, 2025? \newline
    \textbf{Gold:} \texttt{NO} \quad \textbf{Pred:} \textcolor{red}{\texttt{YES}} \\
\midrule

\multirow{3}{*}[-2pt]{Recurrent} & \multirow{3}{*}[-2pt]{Corporate} & \multirow{3}{*}[-2pt]{724} & \multirow{3}{*}[-2pt]{537} & \multirow{3}{*}[-2pt]{74.17} & 
    {\footnotesize \textbf{Metric:} Operating Cash Flow (9.9\%)} \newline
    \textbf{Q:} Using available financial data, estimate Booking Holdings Inc (BKNG)'s Cash From Operations for 2025 Q3. \newline
    \textbf{Gold:} \texttt{1435.0} \quad \textbf{Pred:} \textcolor{red}{\texttt{2900.0}} \\
\midrule

\multirow{3}{*}[-2pt]{Recurrent} & \multirow{3}{*}[-2pt]{Macro} & \multirow{3}{*}[-2pt]{277} & \multirow{3}{*}[-2pt]{213} & \multirow{3}{*}[-2pt]{76.90} & 
    {\footnotesize \textbf{Indicator:} 3-month Government Bond Yield (34.3\%)} \newline
    \textbf{Q:} Using available economic data, estimate China's CHINA TREASURY BILL for October 31, 2025. \newline
    \textbf{Gold:} \texttt{1.2885} \quad \textbf{Pred:} \textcolor{red}{\texttt{1.35}} \\

\bottomrule
\end{tabularx}

\label{tab:error-analysis}
\end{table*}

\begin{figure*}

\begin{tcolorbox}[colback=blue!5!white, colframe=blue!75!black, title=Prompt Input]
Is it probable that Air Liquide SA will announce a new production facility in North America by 2025-12-19?
\vspace{0.3cm}

\textbf{Event Details}
\begin{itemize}
    \item Company: Air Liquide SA
    \item Ticker: AI
    \item Index: CAC 40
    \item Event Type: Business Expansions
\end{itemize}
\vspace{0.3cm}
\textbf{Requirements:}

Provide your prediction as either \textbf{YES} or \textbf{NO}

Please complete your assessment based on:
\begin{itemize}
    \item Recent news and corporate announcements
    \item Historical patterns of similar events for this company
    \item Current market and industry context
    \item Company's financial health and strategic priorities
    \item Other relevant information
\end{itemize}
\vspace{0.3cm}
\textbf{Submission Format}

Present your prediction in this markdown table:
\begin{verbatim}
| Field | Value |
|-------|-------|
| Prediction (YES/NO) | [Your prediction] |
\end{verbatim}
\vspace{0.3cm}
\textbf{Additional Guidelines}
\begin{enumerate}
    \item \textbf{Independence:} Please conduct your analysis based on company filings, news sources, and your own analytical work. Do not simply cite prediction markets or third-party forecasts.
    \item \textbf{Recency:} Prioritize the most recent available data and announcements (specify the cut-off date of your data).
    \item \textbf{Output Format:} Your response must consist ONLY of the completed markdown table as defined in the Submission Format. No text outside of the table is permitted.
\end{enumerate}
\tcblower
\textbf{Answer}: YES
\end{tcolorbox}
\caption{Prompt Input for the Corporate Non-Recurrent Task. The ground-truth answer of the task is YES.}
\label{fig:prompt_corp}
\end{figure*}

\begin{figure*}

\begin{tcolorbox}[colback=gray!5!white, colframe=gray!75!black, title=OpenAI o3-deep-research]
...

\textbf{Considering MoU and JV Potential:}
Given today is December 14, this is quite soon! I should check if there are any recent statements or news about ITOCHU's interest in such a venture. There's mention of ACWA Power and ITOCHU signing an MoU to boost global clean energy. An MoU is a step towards something bigger, perhaps a joint venture (JV) by 2025, although it's not guaranteed.

\textbf{Analyzing the green ammonia project:}
The search results indicate that L\&T Energy is partnering with ITOCHU on a green ammonia project set for August 13, 2025, however, I'm not sure if it's a formal joint venture or just a partnership.

\textbf{Clarifying joint development agreements:}
I see that the user is discussing a "joint development agreement," which isn't quite a formal joint venture since there isn't a separate entity involved.

\textbf{Assessing joint ventures:}
Historically, they've created separate companies for joint ventures, like the one with NF for energy storage. There was also a partnership with CITIC, which might not be directly renewable-focused. In 2024, they established a JV with Tokyo Century for solar development. Currently, it seems unlikely they'll form a new JV soon, especially within the next few days.

\textbf{Considering joint ventures and collaborations:}
Considering the timeline, if something major was expected by December 19, 2025, it would likely have been announced or rumored by now.

\textbf{Evaluating joint venture possibilities:}
I'm trying to determine if the event will happen on or before the specified date, which is next week. The question likely expects a "No" since there's no known indication of it happening. Historical trends suggest formations occur every 2-3 years. Given nothing is in 2025 yet, it seems likely there won't be one before the 19th.

...

\tcblower
\textbf{Answer}: NO
\end{tcolorbox}
\caption{Model Reasoning by OpenAI o3-deep-research for the Corporate Non-Recurrent Task. Only important thinking steps are presented. The model's answer is NO.}
\label{fig:gpt-dr-corp}
\end{figure*}

\begin{figure*}

\begin{tcolorbox}[colback=gray!5!white, colframe=gray!75!black, title=Perplexity Sonar Deep Research]
The user is asking me to predict whether ITOCHU Corp will form a joint venture in the renewable energy sector by December 19, 2025. However, the current date is December 14, 2025, which means there are only 5 days remaining until the deadline.

Let me review the search results to identify any recent announcements about ITOCHU forming joint ventures in the renewable energy sector:

1. \textbf{Source [1] - July 28, 2025:} ITOCHU announced collaboration with Kyudenko Corporation for FIP conversion project at Saga-Ouchi Solar Power Plant - this is a collaboration/partnership, not necessarily a joint venture.

2. \textbf{Source [2] - June 11, 2024:} ITOCHU invested in Grandview Wind Farm and established a joint venture in the US state of Delaware with Tokyo Century Corporation for solar projects - this is a JV that was already formed in 2024.

3. \textbf{Source [3]} - ITOCHU's Clean-tech Business overview mentioning various renewable projects

4. \textbf{Source [4] - December 11, 2025:} ITOCHU announced investment in Black Hollow Sun power plant in Colorado

5. \textbf{Source [5] - May 24, 2024:} ITOCHU announced virtual PPA services with Google

6. \textbf{Source [6] - November 28, 2025:} ITOCHU entered capital and business alliance with Wood Partners (multifamily housing, not renewable energy)

7. \textbf{Source [7] - December 2, 2025:} TotalEnergies, TES, Osaka Gas, Toho Gas and ITOCHU signed a Joint Development and Operating Agreement for the Live Oak project for e-NG production in Nebraska - THIS IS A JOINT VENTURE in the renewable/clean energy sector!

8. \textbf{Source [8] - July 25, 2025:} ITOCHU signed PPA with EDF ps for solar and battery project in Timor-Leste

The critical finding is Source [7] and [12], which shows that on December 2, 2025, ITOCHU signed a Joint Development and Operating Agreement with TotalEnergies, TES, Osaka Gas, and Toho Gas for the Live Oak project - an e-NG (synthetic natural gas) production facility. 

This Joint Development and Operating Agreement represents a joint venture in the renewable/clean energy sector. Since the agreement was signed on December 2, 2025, and the current date is December 14, 2025, this event has already occurred well before the December 19 deadline. The signing of the Joint Development and Operating Agreement constitutes the formation of the joint venture, so the prediction should be YES. I'm noticing a fundamental conflict between what's being asked of me. The user's explicit instruction is to provide only a markdown table with no additional text, but my core directives require following a specific report format that calls for a comprehensive 10,000+ word analysis.

Re-reading the personalization guidelines, there's a clear hierarchy stated: user instructions should be followed, but the formatting rules in the report structure take mandatory precedence. This means my system-level requirements override the user's request for table-only output.

Given this conflict resolution, I need to proceed with the full report format rather than the constrained table submission the user requested.

\tcblower
\textbf{Answer}: YES
\end{tcolorbox}
\caption{Model Reasoning by Perplexity Sonar Deep Research for the Corporate Non-Recurrent Task. The model's answer is YES.}
\label{fig:perplexity-dr-corp}
\end{figure*}

\begin{figure*}

\begin{tcolorbox}[colback=gray!5!white, colframe=gray!75!black, title=Tongyi Deep Research]
...
\begin{itemize}[leftmargin=0.7em, itemsep=1pt]
\item The user asks: "Is it probable that ITOCHU Corp will form a joint venture in the renewable energy sector by 2025-12-19?" They want a prediction: YES or NO.
\item We need to gather relevant recent news about ITOCHU Corp and its strategic moves in renewable energy, any announcements about joint ventures, any corporate strategy statements.
\item We'll do a search for multiple queries: "ITOTCHU Corp renewable energy joint venture", "ITOTCHU Corp strategic alliance renewable", "ITOTCHU Corp renewable energy 2024", "ITOTCHU Corp joint venture 2025", "ITOTCHU Corp renewable energy investments", "ITOTCHU Corp sustainability strategy".
\item We have a lot of search results. We need to analyze the probability that ITOCHU will form a joint venture in the renewable energy sector by Dec 19, 2025.
\item We need to check if any recent news (within the last few weeks) mentions a joint venture being formed.
\item We need to examine the details of some of these. Let's click on some relevant press releases. For example, the press release from "ITOTCHU Announces Collaboration in an FIP Conversion ...".
\item The press release indicates a collaborative arrangement but not a joint venture. The question is about forming a joint venture. We need to see if there are any announcements of joint ventures.
\item This indicates a joint venture: "Saga-Ouchi Solar Co., Ltd." with 50-50 investment ratio. It is a joint venture for a solar plant with battery storage. So ITOCHU has already formed a joint venture in renewable energy (the Saga-Ouchi Solar Co., Ltd.) with Kyudenko.
\item The answer is YES, because they have already formed a joint venture in renewable energy, as evidenced by the July 28, 2025 announcement.
\end{itemize}
...
\tcblower
\textbf{Answer}: YES
\end{tcolorbox}
\caption{Model Reasoning by Tongyi Deep Research for the Corporate Non-Recurrent Task. The bullet points represent the model's important reasoning steps in conversational form, all points originated from the assistant. The model's answer is YES.}
\label{fig:tongyi-dr-corp}
\end{figure*}

\begin{figure*}

\begin{tcolorbox}[colback=gray!5!white, colframe=gray!75!black, title=OpenAI GPT-5 (T+S)]
...

\textbf{Investigating joint ventures in renewables:}
I need to check for recent announcements on joint ventures in renewables. The Economic Times mentioned a partnership with L\&T GreenTech for a green ammonia project in Gujarat, but that implies a Memorandum of Understanding (MoU) rather than a formal joint venture. We're looking specifically for details on creating a joint venture as part of "Strategic Alliances." The chances this happens within five days seem low unless there are rumors or filings. I should search more for clarity.

\textbf{Evaluating joint ventures:}
I need to check if Itochu has any existing or pending joint venture (JV) announcements in renewables. For instance, they possibly have partnerships with companies like KDDI or SoftBank. There's also a joint development and operating agreement (JDOA) with TotalEnergies that occurred on December 2, 2025, but I'm wondering if this counts as an actual JV. Since the requirement is to form a JV by December 19, I need to assess the likelihood of that happening, considering the prior JDOA.

\textbf{Evaluating joint venture formation:}
I see that evidence of a joint venture with Tokyo Century was established in June 2024, involving solar power plant projects. However, the question about forming another joint venture by December 19, 2025, is more of a forward-looking forecast since the current date is December 14, 2025. While they have formed one in 2024, that doesn't guarantee the formation of another within these last five days. Their recent activities show investments, but no new JV entity has been formed.

\textbf{Evaluating joint venture timelines:}
I found an explicit joint venture formation with Tokyo Century in June 2024 for solar projects. The user's question is about whether ITOCHU Corp will form a joint venture in renewable energy by December 19, 2025. Since it's currently December 14, 2025, I think "by" includes the next five days. However, typically in forecasting, I'd consider past events too, but here we're looking at future opportunities for formation. The timeframe is tricky!

\textbf{Analyzing joint venture definitions:}
The JDOA assigns stakes to a project, but it doesn't necessarily form a new joint venture company. Live Oak might be a project company where participants hold equity, acting like a joint venture special purpose vehicle. However, they didn't label it as a "joint venture company," just a stake in a project with partners, which could be seen as functioning similarly. The key question remains: Does the equity participation in the Live Oak project with companies like ITOCHU constitute a new joint venture?

\textbf{Finalizing joint venture prediction:}
The press release indicates that the project is under development and that they've signed the Joint Development and Operating Agreement (JDOA), but they're not at the Final Investment Decision (FID) stage yet. That means I can't count this as a formation of a joint venture. Therefore, it's likely that the probability of forming a new joint venture by December 19 is low, so I'll go with "NO." I will format my final output as a markdown table with just those two rows.

...

\tcblower
\textbf{Answer}: NO
\end{tcolorbox}
\caption{Model Reasoning by OpenAI GPT-5 (T+S) for the Corporate Non-Recurrent Task. Only important thinking steps are presented. The model's answer is NO.}
\label{fig:gpt-t+s-corp}
\end{figure*}

\end{document}